\documentclass[10pt,aps,prd,amsmath,amssymb,superscriptaddress,twocolumn,nofootinbib,showpacs,preprintnumbers,amsfonts, notitlepage,floatfix]{revtex4-1}
\usepackage[utf8]{inputenc}
\usepackage{mathtools}

\usepackage[usenames,dvipsnames,svgnames,table]{xcolor} 
\usepackage{graphicx}
\usepackage{dcolumn}
\usepackage{bm}
\usepackage{hyperref}

\usepackage[capitalize]{cleveref}
\usepackage{dsfont}
\usepackage{placeins}
\usepackage{todonotes}
\usepackage[utf8]{inputenc}
\usepackage{multirow}
\usepackage{amsmath}
\usepackage{amsfonts}
\usepackage{amssymb}
\usepackage{enumitem,amssymb}

\usepackage{array}   
\newcolumntype{L}{>{$}l<{$}} 
\newcolumntype{R}{>{$}r<{$}}
\newcolumntype{C}{>{$}c<{$}}
\usepackage{xspace}
\usepackage{braket}
\usepackage{units}
\usepackage{slashed}
\usepackage[normalem]{ulem}
\usepackage[format=plain,justification=RaggedRight,singlelinecheck=false]{caption}
\usepackage[format=plain,justification=centering,singlelinecheck=false]{subcaption}
\usepackage{tabularx}
\usepackage{blindtext}
\usepackage{gensymb}
\usepackage{textcomp}
\usepackage{soul,color}

\usepackage[letterspace=-10]{microtype} 

\definecolor{jlab_red}{RGB}{192,39,45}
\definecolor{jlab_orange}{RGB}{249,102,0}
\definecolor{jlab_blue}{RGB}{47,122,121}
\definecolor{jlab_green}{RGB}{65,125,10}

\hypersetup{%
pdftitle = {arxiv_v2},
pdfsubject = {QCD},
pdfkeywords = {QCD, Hadron, Physics, Lattice, Meson, Scattering},
pdfauthor = {Hadron Spectrum Collaboration},
colorlinks = {true},
filecolor = {black},
linkcolor = {jlab_blue},
menucolor = {black},
citecolor = {jlab_green},
urlcolor = {jlab_green},
}{}

\allowdisplaybreaks

\begin{document}

\preprint{JLAB-THY-23-3778}

\title{Quark mass dependence of $\pi \pi$ scattering in isospin 0, 1, and 2 from lattice QCD}

\author{Arkaitz~Rodas}
\email{arodas@jlab.org}
\affiliation{\lsstyle Thomas Jefferson National Accelerator Facility, 12000 Jefferson Avenue, Newport News, VA 23606, USA}
\author{Jozef~J.~Dudek}
\email{dudek@jlab.org}
\affiliation{\lsstyle Thomas Jefferson National Accelerator Facility, 12000 Jefferson Avenue, Newport News, VA 23606, USA}
\affiliation{Department of Physics, College of William and Mary, Williamsburg, VA 23187, USA}
\author{Robert~G.~Edwards}
\email{edwards@jlab.org}
\affiliation{\lsstyle Thomas Jefferson National Accelerator Facility, 12000 Jefferson Avenue, Newport News, VA 23606, USA}

\collaboration{for the Hadron Spectrum Collaboration}
\date{\today}

\begin{abstract}
Using lattice QCD we extract $\pi\pi$ scattering amplitudes with isospin--0,1,2 in low partial-waves at two values of the light quark mass corresponding to $m_\pi \sim 283$ and $330$ MeV. We confirm expectations of weak repulsion in isospin--2, and the presence of a narrow $\rho$ resonance in isospin--1, and study the pion mass dependence of these channels. In isospin--0 we find that the two pion masses considered straddle the point at which the $\sigma$ transitions from being a stable bound-state to being either a virtual bound-state or a subthreshold resonance. We discuss the ability of lattice calculations like these to precisely determine the $\sigma$ pole location when it is a resonance, and propose an approach in which the full complement of amplitudes computed in this paper can be used simultaneously to provide more constraint.
\end{abstract}
\maketitle

\section{Introduction}
 \label{sec:introduction}
 
Hadron-hadron scattering processes have long been used as a tool to explore strong interaction physics. The amplitudes which describe these processes as a function of energy and angle can be expanded in partial-waves, and examination of these yields information about the resonance content of quantum chromodynamics. Scattering of the lightest hadron, the pion, off the pion cloud around a proton or nucleus offers the simplest such process, being unburdened by complications of spin.

Experimentally, $\pi\pi$ scattering in the lowest partial-waves shows very different behavior across the three possible isospins. Isospin--2 is found to be weak and repulsive, with the lack of resonances being an early motivator of the $q\bar{q}$ quark model. Isospin--1 houses the narrow $\rho$ resonance, appearing as a rapid rise in the phase-shift of the $P$--wave amplitude, which is otherwise featureless at low energy. Isospin--0 is found to be attractive, but shows only a slow rise in phase-shift with energy across the elastic scattering region. Typically, this rise is associated with a \emph{broad} scalar resonance, the $\sigma$. A state with these quantum numbers has historically been included in models of the nucleon-nucleon potential to describe intermediate-range effects. Precise determination of the pole location of the $\sigma$ remained a problem until recently, with a range of amplitude parameterizations applied to experimental data generating a spread of pole locations~\cite{ParticleDataGroup:2022pth}. This problem was solved by applying dispersion relations which implement analyticity and crossing symmetry in a consistent way, providing additional constraint beyond that given by the isospin--0 scattering data on the real energy axis alone~\cite{Ananthanarayan:2000ht,Colangelo:2001df,Kaminski:2006yv,Kaminski:2006qe,GarciaMartin:2011cn,Moussallam:2011zg}.

The nature of the $\sigma$ within QCD is somewhat unclear~\cite{Jaffe:1976ig,Maiani:2004uc,tHooft:2008rus,RuizdeElvira:2010cs,Guo:2015daa}. It is often partnered with the $f_0(980),\, a_0(980)$ and $\kappa$ states into a `scalar nonet', despite the very different appearances of these resonances (narrow states at $K\bar{K}$ threshold versus very broad states away from any threshold). An association of this type for the lightest vector resonances, $\rho, \omega, \phi, K^*$, is quite natural, given their common properties, and is often used as motivation for a $q\bar{q}$ quark-model assignment of these states, with their modest differences being due to the mild breaking of an approximate $SU(3)$ flavor symmetry that leads to states with strange content being heavier. The scalar nonet has no such simple interpretation~\cite{Morgan:1974cm,Tornqvist:1982yv,Tornqvist:1995ay,Tornqvist:1995kr}.

Recently, $\pi\pi$ scattering in QCD has been considered making use of the first-principles lattice approach. The discrete spectrum of states with the quantum numbers of $\pi\pi$ in the finite periodic spatial volume of the lattice can be related to $\pi\pi$ scattering amplitudes through the L\"uscher relation~\cite{Luscher:1986pf,Briceno:2017max}. Computations have taken place at several values of the light-quark mass for all three isospins~\cite{
Sharpe:1992pp,Gupta:1993rn,Kuramashi:1993ka,Fukugita:1994ve,JLQCD:2002mgw,Du:2004ib,CP-PACS:2004dtj,Beane:2005rj,Beane:2007xs,CLQCD:2007rcz,Feng:2009ij,Yagi:2011jn,Sasaki:2013vxa,Fu:2013ffa,ETM:2015bzg,Dudek:2010ew,NPLQCD:2011htk,Dudek:2012gj,Bulava:2016mks,Horz:2019rrn,Culver:2019qtx,Fischer:2020jzp,CP-PACS:2007wro, Feng:2010es, Lang:2011mn, CS:2011vqf, Dudek:2012xn, Pelissier:2012pi, Wilson:2015dqa, Bali:2015gji, Bulava:2016mks, Alexandrou:2017mpi,Andersen:2018mau,ExtendedTwistedMass:2019omo,Fischer:2020yvw,Briceno:2016mjc,Fu:2017apw,Guo:2018zss,RBC:2021acc}.
By parameterizing the elastic scattering amplitudes the resonance pole content can be investigated through analytic continuation into the complex energy plane.

Isospin--2 is found to be weak and repulsive, as in experiment, and the variation of the scattering length with changing quark mass has been explored in the context of chiral perturbation theory~\cite{NPLQCD:2011htk,Fischer:2020jzp}. Isospin--1 is found to feature a $\rho$--like resonance whose mass increases and width decreases with increasing quark mass until it becomes stable at a pion mass near 400 MeV. Isospin--0 shows a much more dramatic evolution with changing quark mass: at $m_\pi \sim 391$ MeV, a clear \emph{stable} bound-state $\sigma$ is observed, while at $m_\pi \sim 239$ MeV, a slow variation of the phase-shift appears to indicate a broad resonance $\sigma$, albeit with a large degree of amplitude parameterization dependence in the pole position~\cite{Briceno:2016mjc,Briceno:2017qmb}.

The possibility that the $\sigma$ could undergo a transition from being a broad resonance into being bound, as the light quark mass is increased, was previously explored in a unitarized version of chiral perturbation theory~\cite{Hanhart:2008mx,Guo:2018zss}. By making assumptions about the quark mass dependence of certain low-energy coefficients, it was found that over a relatively small variation in pion mass, the $\sigma$ undergoes a rapid transition from being a bound state, to being a virtual bound state (a pole on the real energy axis below threshold on the unphysical Riemann sheet), to being a broad resonance. These results provide a particular manifestation of the general framework for scalar resonance pole trajectory discussed in Ref.~\cite{Hanhart:2014ssa}.

In this paper, we will report the results of a calculation determining $\pi\pi$ scattering amplitudes in all three isospins at two previously unconsidered light quark masses, corresponding to $m_\pi \sim $ 283 and 330 MeV. These values lie between the points at which the $\sigma$ has been observed in lattice calculations to be bound, and where it appears as a broad resonance, so that we aim to be able to close in on the region where the transition takes place.

The scatter of $\sigma$ pole positions under reasonable variation of amplitude parameterization in the previous lattice calculation at $m_\pi \sim 239$ MeV indicated that the same issue present in analysis of experimental scattering data may plague attempts to pin down with precision the pole location in these first-principles QCD efforts. A possible mechanism to overcome this might be to apply dispersion relations to the results of lattice QCD computations. Such an approach would require input of computed $\pi\pi$ scattering amplitudes in all three isospins in low partial waves, which is what we provide in this paper.

We will show that the isospin--2 amplitude evolves smoothly with changing light quark mass, and we will explore the role of the `Adler zero' predicted by the broken chiral symmetry of QCD. The evolution of the $\rho$ resonance which dominates the isospin--1 amplitude is presented, with a confirmation of the near independence of its coupling to $\pi\pi$ to variations in the light quark mass.   The isospin--0 $S$-wave amplitude is found to undergo a dramatic transition between ${m_\pi \sim  330}$ MeV and 283 MeV, from a behavior compatible with an only-just-bound $\sigma$ at the heavier mass to a mild energy dependence compatible with being either a virtual bound state or a subthreshold resonance at the lighter mass. We will conclude that to make more precise statements about the $\sigma$ in cases that it is unbound, we require additional constraints of the type offered by dispersion relations\footnote{While this paper was in the final stages of production, a preprint, Ref.~\cite{Cao:2023ntr}, appeared applying dispersion relations to lattice QCD data, focusing at $m_\pi \sim 391$ MeV, where the $\sigma$ is a well-determined bound-state.}.   

\section{Lattices and operator construction}
 \label{sec:latt}

\begin{table*}
\begin{tabular}{rrlccc|lllll|c}
 
  $-a_t m_\ell$ & $(L/a_s)^3$ & $\times \; T/a_t$ & $N_{\mathrm{cfgs}}$ & $N_\mathrm{vecs}$ & $N_\mathrm{t_{src}}$   & \multicolumn{1}{c}{$a_t m_\pi$} & \multicolumn{1}{c}{$a_t m_K$} & \multicolumn{1}{c}{$a_t m_\eta$} & \multicolumn{1}{c}{$a_t m_\Omega$} & \multicolumn{1}{c|}{$\xi$} & $m_\pi$/MeV\\[0.2ex]
\hline

 $0.0850$ & $24^3$ &$\times$ 256 & 473 & 160 & 8--16  & 0.05635(14) & 0.09027(15) & 0.09790(100)         & 0.2857(8)  & 3.467(8)     & 330  \\
 $0.0856$ & {\footnotesize {$24^3$,\! $32^3$}} &$\times$ 256 & {\footnotesize 393,\! 475} & {\footnotesize 160,\! 256} & 4--8  & 0.04720(11) & 0.08659(14) & 0.09602(70) & 0.2793(8)  & 3.457(6)     & 283  \\
\end{tabular}
\caption{
Anisotropic three-flavor lattices used in this paper. Anisotropy values, $\xi$, are obtained from the pion dispersion relation. $N_{\text{vecs}}$ indicates the number of distillation vectors used in the construction of correlation functions, and $N_\mathrm{t_{src}}$ the number of $0 \to t$ perambulator time-sources averaged over~\cite{HadronSpectrum:2009krc}.
}
\label{tab:lattices}
\end{table*}

The calculations described in this manuscript make use of anisotropic Clover lattices~\cite{Edwards:2008ja, HadronSpectrum:2008xlg} whose parameters are presented in Table~\ref{tab:lattices}. These three-flavor lattices, which have $a_s \sim 0.12 \, \mathrm{fm}$, degenerate light quarks, and a strange quark mass approximately tuned to the physical strange quark mass, were previously used in calculations of the $\pi K$ system~\cite{Wilson:2019wfr,Radhakrishnan:2022ubg}~\footnote{
The pion masses have been recomputed with greater statistics since that previous paper, and herein are referred to as 283, 330 MeV, which correspond to the 284, 327 MeV lattices therein.
}.

In order to determine $\pi\pi$ scattering amplitudes, we require the spectra of states with the appropriate quantum numbers in the finite spatial volume of the lattice. These spectra are extracted using variational analysis of matrices of two-point correlation functions computed using a basis of operators at source and sink. A basis that has proven successful in prior calculations~\cite{Dudek:2012gj,Dudek:2012xn,Dudek:2014qha,Wilson:2014cna,Briceno:2016mjc,Briceno:2017qmb} makes use of `single-hadron' operators (in isospin--0 and isospin--1) of fermion bilinear type, $\bar{\psi} \Gamma \overleftrightarrow{D} \ldots \overleftrightarrow{D} \psi$, supplemented by operators resembling a pair of mesons having definite total momentum, and magnitude of relative momentum,
\begin{equation}
	\big( \Omega\,\Omega \big)_{\vec{P}, \Lambda,\mu}^{\dag\,\,[\vec{p}_1, \vec{p}_2]} = \sum
	_{\substack{ \vec{p}_1 + \vec{p}_2 = \vec{P} }}  \mathcal{C}(\vec{P},\Lambda,\mu; \vec{p}_1; \vec{p}_2 )\, \Omega^\dag(\vec{p}_1)\, \Omega^\dag(\vec{p}_2).
\end{equation} 
The operators appearing in the product on the right-hand side are selected to be those linear combinations of `single-hadron' operators that optimally overlap with the pion states in the variational analysis of correlation functions with the quantum numbers of a single pion. The `lattice Clebsch-Gordan coefficients' in this equation ensure that the operator transforms in a definite irreducible representation, $\Lambda$, of the relevant lattice symmetry group. Systems of definite angular momentum subduce into these `irreps' according to Table~II of Ref.~\cite{Dudek:2012gj} for $I=0,\,2$, and according to Table~III of Ref.~\cite{Dudek:2012xn} for $I=1$. In this paper, we will consider irreps with total momentum $|\vec{P}|^2 \le 4 \left(2\pi/L\right)^2$.

Use of the distillation framework~\cite{Peardon:2009gh} allows for efficient computation of a large number of correlation functions, and in particular, allows diagrams featuring quark-antiquark annihilation (of which there are many in the isospin--0 case) to be evaluated without further approximation~\cite{Briceno:2016mjc,Briceno:2017qmb}.

While our primary focus is on $\pi \pi$ elastic scattering, in order to have a reliable evaluation of the finite-volume spectra in the energy region where the $K\bar{K}$ and $\eta\eta$ thresholds open, we have included, where relevant, also $K\bar{K}$--like and $\eta\eta$--like operators into our basis. We are guided as to which relative momentum combinations to include in the basis by the \emph{non-interacting} energy of the operator, $E_\mathrm{n.i.} = \sqrt{ m^2 + |\vec{p}_1|^2} + \sqrt{ m^2 + |\vec{p}_2|^2}$, where $m$ is the mass of the meson ($\pi, K, \eta$). All operator constructions are included which have non-interacting energy in the energy region we intend to study.

The matrices of correlation functions computed in the large basis indicated above are analyzed using a variational approach based upon solving a generalized eigenvalue problem. Our primary interest is in the spectrum which is obtained by fitting the exponential time-dependence of the extracted eigenvalues. In order to account for the impact of the choice of fitting window and the number of exponentials, we implement a version of the ``model averaging" technique proposed in Ref.~\cite{Jay:2020jkz}, as described in Ref.~\cite{Radhakrishnan:2022ubg}. In addition, the sensitivity of the extracted energy levels to the choice of the reference timeslice $t_0$ in the generalized eigenvalue problem, and to the precise choice of operators in the basis is explored and reflected in the quoted energy values and uncertainties.

When dimensionful quantities are required, the lattice scale is set using the $\Omega$ baryon mass computed on the relevant lattice, $a_t=\frac{a_t m_{\Omega}}{m_{\Omega}^{\text {phys }}}$, where the physical mass of the $\Omega$ baryon is $m_{\Omega}^{\text {phys }}=1672.45$ MeV. The quoted pion mass in MeV for each lattice follows from use of this prescription.

\section{Extracting scattering amplitudes from finite-volume spectra}
 \label{sec:amplitude-analysis}

The relationship between two-body scattering amplitudes and the discrete spectrum of states in a finite periodic volume is well established~\cite{Luscher:1986pf, Luscher:1990ux, Luscher:1991cf, Rummukainen:1995vs, He:2005ey, Christ:2005gi, Kim:2005gf, Guo:2012hv, Hansen:2012tf, Briceno:2012yi, Briceno:2014oea}. For an irrep $\Lambda$ of total momentum $\vec{P}$, the discrete spectrum in an $L \times L \times L$ box corresponds to the solutions of
 \begin{equation}
\det \left[ \mathbf{1} + i \boldsymbol{\rho}(E) \cdot \mathbf{t}(E) \cdot \Big( \mathbf{1} + i \boldsymbol{\mathcal{M}}^{\vec{P}, \Lambda}(E,L)   \Big)\right] = 0 \,,
\label{eq:qcond}
\end{equation}
where $\boldsymbol{\mathcal{M}}(E,L)$ is a matrix of known kinematic functions which characterize the cubic spatial volume, while $\mathbf{t}(E)$ contains the partial-wave scattering amplitudes. In general, these are matrices in the spaces of coupled-channels and partial-wave angular momentum, $\ell$, but for elastic scattering they reduce to being dense and diagonal matrices respectively in $\ell$.

Our approach follows from parameterizing the energy-dependence of the partial-wave amplitudes $t^I_\ell(s)$ for those lowest values of $\ell$ which subduce into the irrep $\vec{P}, \Lambda$. In practice for $I=1$, only $\ell=1$ is relevant over the elastic region, while for $I=0,2$, both $\ell=0$ and $\ell=2$ are considered. For a given set of parameter values in these parameterizations, the solution of~\cref{eq:qcond} yields a discrete spectrum that can be compared to the lattice QCD computed spectrum via a correlated $\chi^2$. We form this $\chi^2$ by considering energy levels in all irreps which constrain the partial-waves for each choice of isospin, and take as the amplitude results those which minimize the $\chi^2$. Explicit expressions for the subduction of partial-waves into irreps of the relevant symmetry group are presented in Refs.~\cite{Dudek:2016cru} and \cite{Thomas:2011rh}, and further discussion of the approach, and implementation details can be found in Refs.~\cite{Briceno:2017max, woss:2020cmp, Dudek:2016cru,Briceno:2017qmb}.

The elastic scattering partial-wave amplitudes appearing in~\cref{eq:qcond} can be parameterized by compact forms, allowing for a description of the entire lattice QCD computed spectrum in terms of a few free parameters. The resulting amplitudes can be analytically continued into the complex energy plane to search for pole singularities. A range of parameterization forms is typically used, with the spread of amplitude behaviors and pole locations providing an estimate of systematic error. Each relevant partial-wave amplitude $t^I_\ell(s)$ is parameterized in a way that respects elastic unitarity exactly, but may not necessarily respect other fundamental constraints. 

In terms of the \emph{phase-shift}, $\delta_{\ell}^{I}(s)$, elastic amplitudes can be written as
\begin{equation}
t_{\ell}^{I}(s)=\frac{1}{\rho(s)} \, e^{i \delta^{I}_{\ell}(s)} \sin \delta_{\ell}^{I}(s)=\frac{1}{\rho(s)} \frac{1}{\cot \delta_{\ell}^{I}(s)-i},
\label{eq:tamp}
\end{equation}
where $\rho(s) = 2 k/\sqrt{s}$ is the $\pi \pi$ phase-space, with ${k = \tfrac{1}{2}\sqrt{ s - 4m_\pi^2}}$ being the scattering momentum. 

In those cases where a single partial-wave only dominates~\cref{eq:qcond}, or where the amplitudes for higher partial waves are fixed, each discrete finite-volume energy can be used to obtain a discrete value of the phase-shift at that energy. This approach is used to make the discrete phase-shift `data points' that will appear in plots later in this document. The amplitude curves are not obtained by fitting these `data', but rather using the spectrum $\chi^2$ approach described above.

At low scattering energies, the slow variation of the $S$--wave and $D$--wave can be well described by a low-order expansion in the square of the scattering momentum, typically called the \emph{effective range expansion},
\begin{equation}
k^{2\ell+1} \cot \delta_{\ell}^{I} =  F^I_{\ell}(s)\left( \tfrac{1}{a^I_\ell}+\tfrac{1}{2}r^I_\ell k^2+\dots\right),
\label{eq:cotere}
\end{equation}
where the conventional choice has $F^I_{\ell}(s) = 1$, $a_\ell^I$ interpreted as the \emph{scattering length} and  $r_\ell^I$ as the \emph{effective range}. Additional desired features can be built into the amplitude with other choices, such as $F^I_\ell(s)= \frac{4m_\pi^2 - s_A}{s-s_A}$ to ensure a zero of the amplitude, like those predicted by broken chiral symmetry known as `Adler zeroes'.

An alternative expansion follows from defining
\begin{equation}
\Phi^I_{\ell}(s) = \tfrac{2}{\sqrt{s}} k^{2 \ell+1} \cot \delta^I_{\ell}(s),
\end{equation}
which must be real analytic between the elastic threshold and the inelastic threshold. One can engineer the presence of an effective inelastic threshold ($s_0$), and the opening of the \emph{left-hand-cut} at $s = 0$ (required by crossing-symmetry), by using a \emph{conformal mapping variable}~\cite{Cherry:2000ut,Pelaez:2016tgi},
\begin{equation}
\omega(s) = \frac{\sqrt{s}-\alpha\sqrt{s_0-s}}{\sqrt{s}+\alpha\sqrt{s_0-s}} \, .
\end{equation}
In this expression $\alpha$ and $s_0$ are fixed parameters that determine what energy region is mapped into a unit disk of $\omega$~\footnote{In practice, we will set $s_0 = 0.09 \,a_t^{-2}$ and $\alpha=0.8$ for the $I=2$ waves  and the $I=0$ $D$--wave, as they do not exhibit any inelastic behavior up to high energies. We use $s_0 = 0.05 \,a_t^{-2}$ and $\alpha=1$ for the $I=1$ $P$--wave. For the $I=0$ $S$--wave, where we expect the inelasticity to become significant at a lower energy, we set $\alpha=1$, and we consider two values of $s_0 = 0.032 \,a_t^{-2},\, 0.04 \,a_t^{-2}$.}. The convergence of the conformal expansion is expected to be rapid,
\begin{equation}
\Phi^I_\ell(s)= F^I_{\ell}(s) \sum_{n=0}^N B_{n} \, \omega^{n} \, ,
\end{equation}
where, again, one may build additional properties into the amplitude by suitable choices for $F^I_{\ell}(s)$, for example, $F^I_\ell(s) = \frac{s-m^2_R}{m_R^2}$, to force a resonance. As suggested in Ref.~\cite{Yndurain:2007qm}, spurious singularities introduced below threshold by this conformal expansion can be removed by adding a suitable function,
\begin{equation}
\Phi^I_\ell(s)= F^I_{\ell}(s) \left(\gamma^I_\ell(s)+\sum_{n=0}^N B_{n} \, \omega^{n} \right)\, .
\end{equation}

Partial-waves that contain a narrow resonance and no other features, like the experimental $I=1$ $P$--wave, are usually well-described over a limited energy region by a Breit-Wigner form, which effectively parameterizes a single nearby pole,
\begin{equation}
t_{\ell=1}(s)=\frac{1}{\rho(s)} \frac{\sqrt{s}\,  \Gamma(s)}{m_\mathrm{BW}^2 - s - i \sqrt{s}\,  \Gamma(s)} \, ,
\label{eq:bw}
\end{equation}
with the energy-dependent width, $\Gamma(s) = \tfrac{g_\mathrm{BW}^2}{6\pi} \tfrac{k^3}{s}$. The width form can be elaborated to damp out the threshold behavior at high energies (sometimes called barrier factors) at the cost of including at least one extra parameter and possibly spurious singularities.

A rather flexible parameterization scheme which generalizes nicely to the case of \emph{coupled-channel} amplitudes, uses the $K$-matrix defined in~\cite{Aitchison:1972ay}
\begin{equation}
\left( t^I_\ell(s) \right)^{-1} = \left( K^I_\ell(s) \right)^{-1} - i \rho(s) \, ,
\end{equation}
where a common parameterization choice is a sum of poles plus a finite-order polynomial,
\begin{equation}
K^I_\ell(s) = (2k)^{2\ell} \left[ \sum_r \frac{g_r^2}{m_r^2 - s} + \sum_p \gamma_p s^p \right] \, .
\end{equation}
This form can be modified to ensure an Adler zero by taking $K(s) \to (s-s_A)\,  K(s)$, and the unphysical singularity in the phase-space at $s=0$ can be removed from the physical sheet by replacing $-i\rho(s)$ with the Chew-Mandelstam function, which we present subtracted at threshold, as
\begin{equation}
I(s)=\frac{\rho(s)}{\pi} \log \left[\frac{\rho(s) +1}{\rho(s) - 1}\right] \, ,
\label{eq:chewman}
\end{equation}
which has $\mathrm{Im} \, I(s) = - \rho(s)$ as required by unitarity. 

For each partial wave, we will consider a large number of parameterizations based on the forms above, reporting all those which are found to be capable of describing the computed finite-volume spectra as established by the spectrum $\chi^2$ value.

For every amplitude parameterization, once the parameters are constrained by describing the lattice QCD spectra, we search the second Riemann sheet for poles that we interpret as being due to resonances. The pole locations provide a model-independent definition of a mass and width for the resonance, $\sqrt{s_R} = m_R-i\,\Gamma_R/2$, and the corresponding residue in $t^I_\ell(s) \sim c^2/(s_R - s )$, gives a coupling of the resonance to $\pi\pi$. An alternative definition of the $\pi\pi$ coupling, as presented in Ref.~\cite{Garcia-Martin:2011nna}, is related to ours by,
\begin{equation}
g_{\pi\pi}^2 = 16\pi \frac{2\ell+1}{(2k_R)^{2\ell}} \, c^2 . \label{gNorm}
\end{equation}

We will find, as has been observed in previous lattice calculations~\cite{Dudek:2012xn, Wilson:2015dqa, Briceno:2017qmb, Wilson:2019wfr, Woss:2019hse, Johnson:2020ilc}, and in amplitude analyses of experimental data~\cite{Au:1986vs,JPAC:2017dbi,JPAC:2018zyd,Rodas:2021tyb}, that when a narrow resonance is present the pole position and coupling typically show very little scatter over a range of sufficiently flexible parameterizations, but when a resonance pole lies far into the complex plane, different amplitudes which behave similarly in a limited energy region on the real energy axis (and which describe the finite-volume spectra equally well in the lattice case) can lead to quite different pole locations~\cite{Pelaez:2015qba,Ropertz:2018stk,Pelaez:2020gnd}. We will return to this point later when discussing the $\sigma$ pole appearing in the isospin--0 $S$--wave.

In the following, we will illustrate the finite-volume energy levels included in our fits in black, to discern them from other levels that are not fitted, plotted in gray.

\section{$\pi \pi \to \pi \pi$ $I=2$}
 \label{sec:I2pipi}

%
\begin{figure*}[htbp]
\begin{center}
\resizebox{\textwidth}{!}{
  \includegraphics{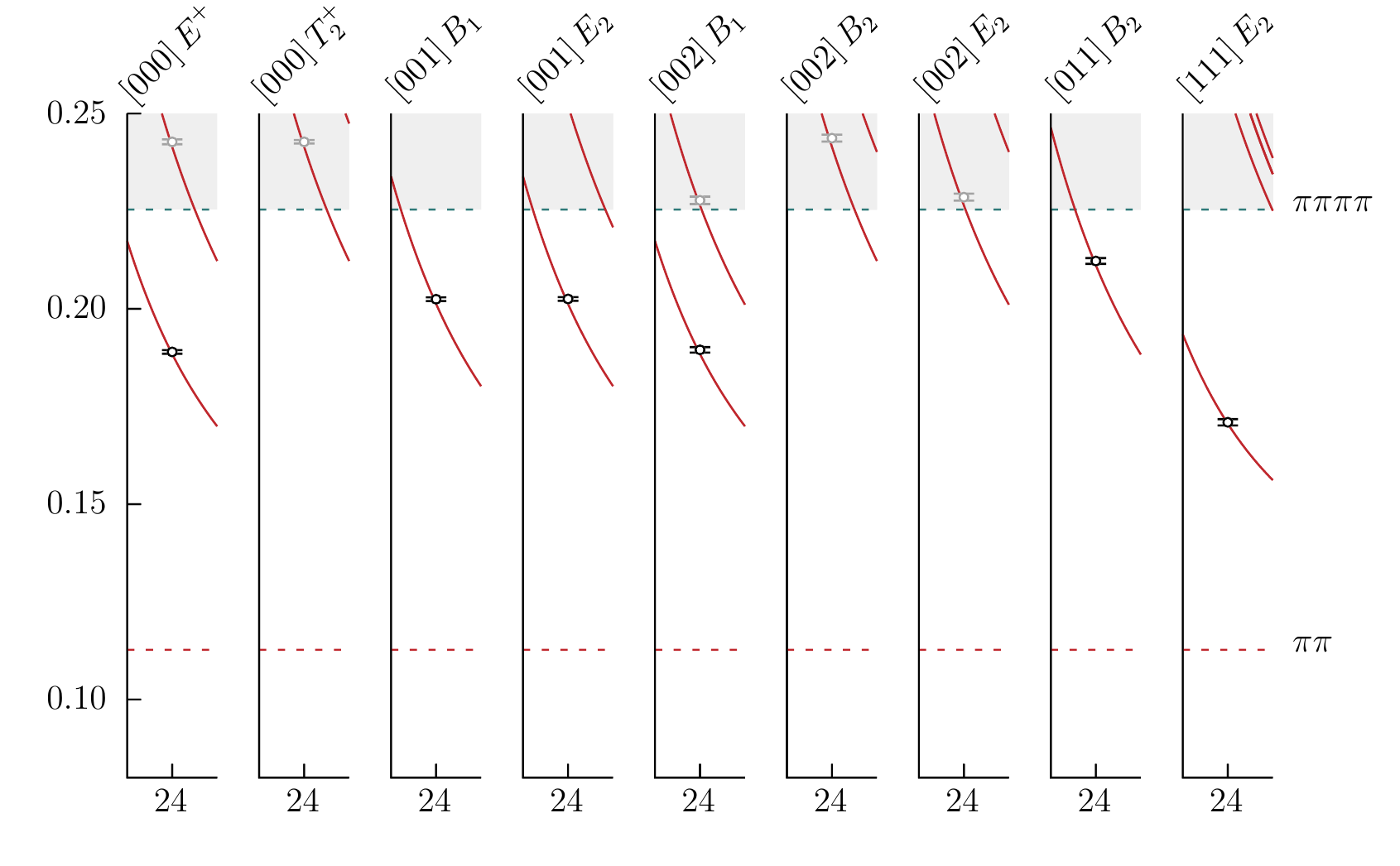}
  \includegraphics{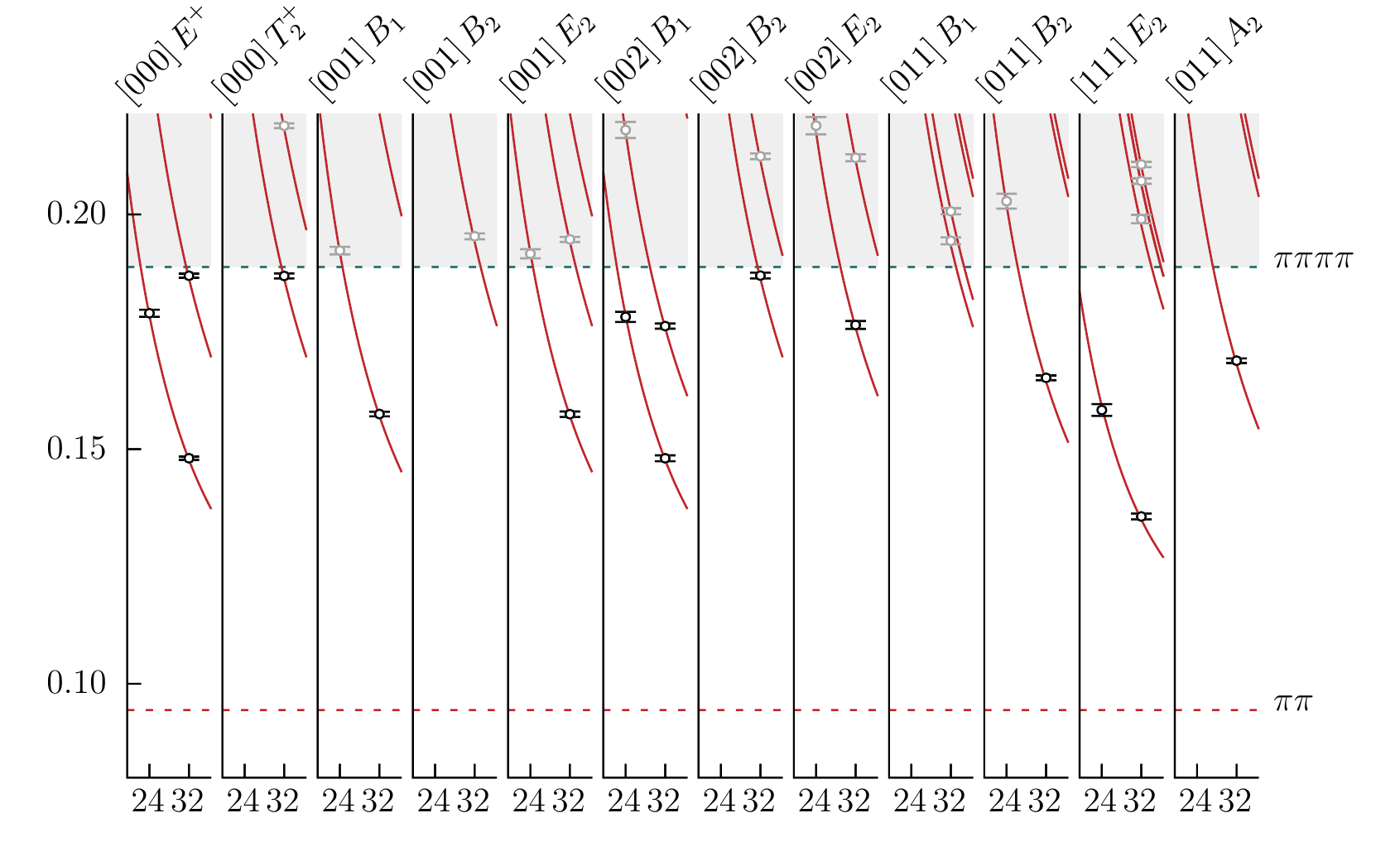}
}
\resizebox{\textwidth}{!}{
  \includegraphics{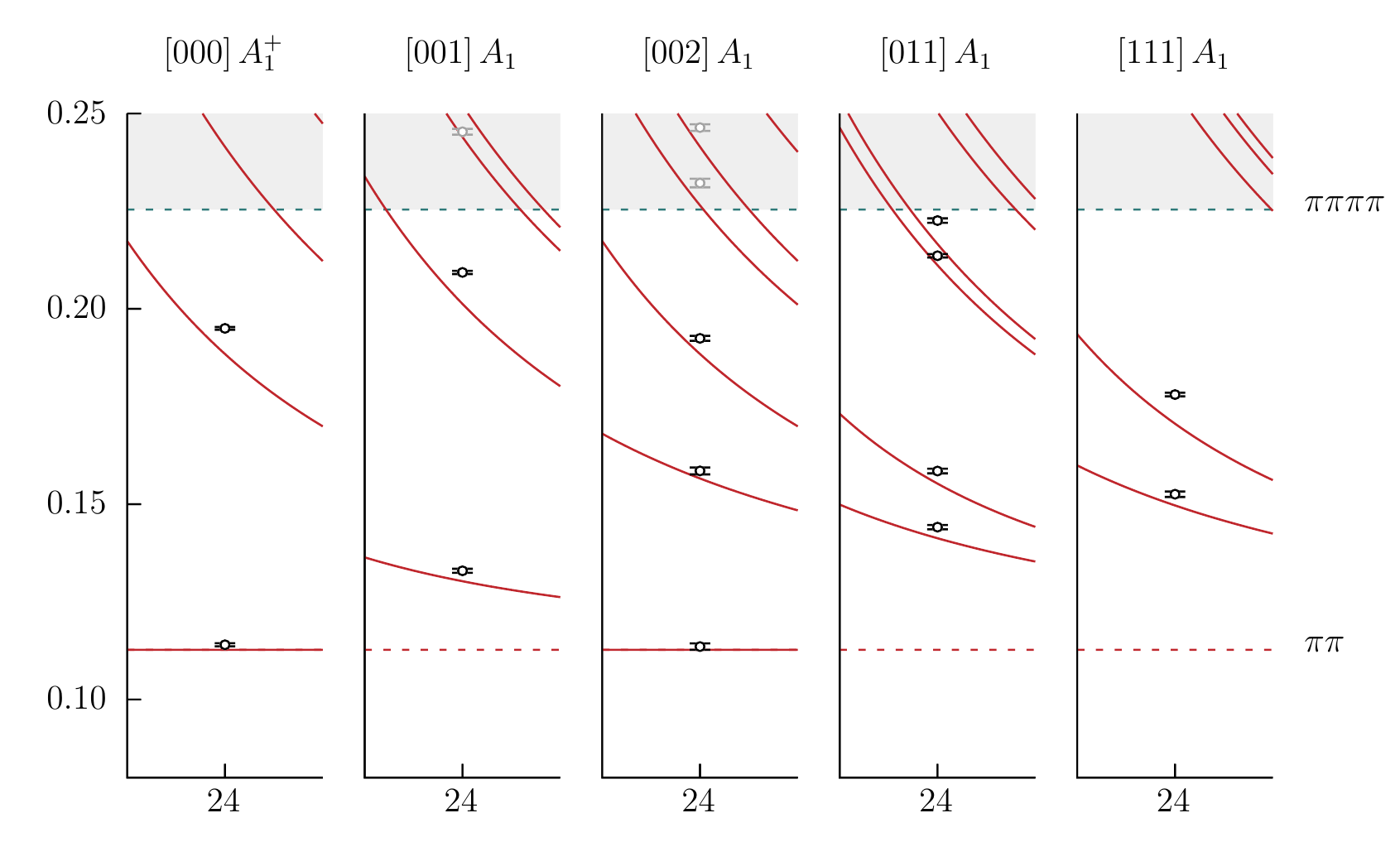}
  \includegraphics{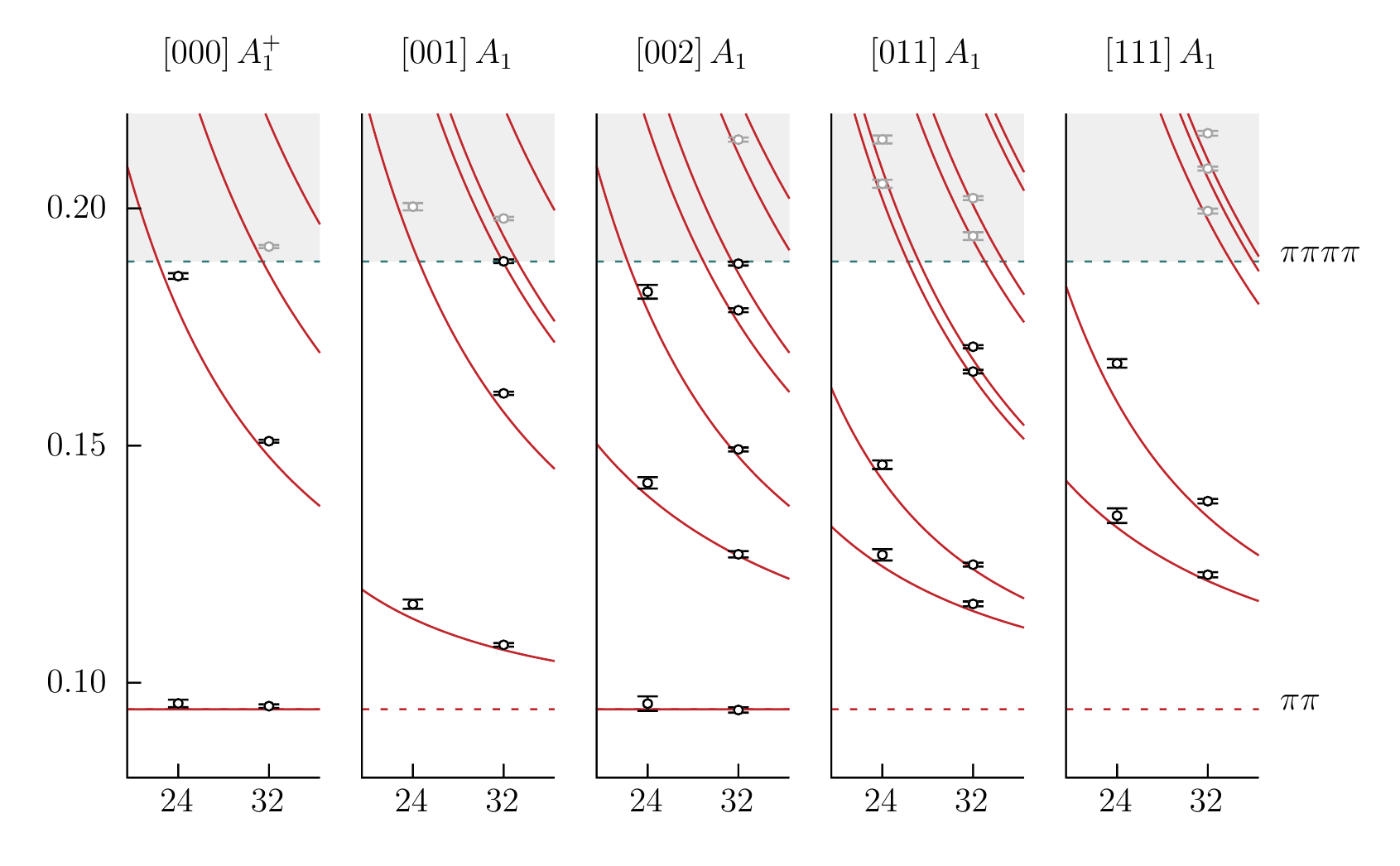}
}
\caption{
$I=2$ finite-volume spectra, $a_t E_\mathsf{cm}$, by irrep, against $L/a_s$, for $m_\pi \sim 330$ MeV (left) and $m_\pi \sim 283$ MeV (right). Upper panel irreps have $D$-wave as their leading partial-wave, while those in the lower panel have $S$-wave leading. Red curves show $\pi\pi$ non-interacting energies.
}
\label{fig:I2}   
\end{center}
\end{figure*}

As in experiment, previous determinations in lattice QCD at various pion masses (e.g. ~\cite{Beane:2007xs,Dudek:2010ew,NPLQCD:2011htk,Dudek:2012gj,Kurth:2013tua,Fu:2013ffa,Bulava:2016mks}) have found $\pi\pi$ scattering in isospin--2 to be weak and repulsive. Lattice calculations of this channel typically use a basis of operators resembling a pair of pions only, since $q\bar{q}$ operators cannot access $I=2$. The lowest inelastic channel is $\pi\pi\pi\pi$, but expectations from experiment are that the coupling of this channel to $\pi\pi$ turns on very slowly~\cite{Losty:1973et,Durusoy:1973aj,Hoogland:1977kt}.

\subsection{$I=2$ finite-volume spectra}
\label{subsec:I2FV}

Using bases of $\pi\pi$ operators as described in Section~\ref{sec:latt}, matrices of correlation functions are computed, and variational analysis leads to the spectra shown in~\cref{fig:I2}. Departure of the discrete energy levels from the values for non-interacting $\pi\pi$ pairs can be observed, being much larger in those irreps which feature a subduction of the $S$--wave.

The first inelastic threshold here is $\pi\pi\pi\pi$, indicated in~\cref{fig:I2} by the horizontal dashed line. We have not included any $\pi\pi\pi\pi$-like operator constructions in our basis, so the spectrum presented above the inelastic threshold will only be a correct subset of the complete true spectrum in the case that the $\pi\pi$ and $\pi\pi\pi\pi$ channels are decoupled. In experiment, this is indeed the case until quite high energies, well above those considered here~\cite{Losty:1973et,Durusoy:1973aj,Hoogland:1977kt}.

The determined energies have fractional errors typically at the 0.5\% level, where this includes an estimate of systematic error coming from varying different fitting details and whether a ``weighting-shifting'' step (see Ref.~\cite{Dudek:2012gj}) is applied to cancel mild finite-time-extent effects. These systematic variations impact at a level below the statistical error on most points. Across all irreps, we extract 50 energy levels for $m_\pi \sim 330$ MeV, and 98 for $m_\pi \sim 283$ MeV, of which 19 and 41 are below or at the $\pi\pi\pi\pi$ threshold, respectively.

%
%
\begin{figure}[htbp]
  \includegraphics[trim=0 6 0 3, width=\columnwidth, clip]{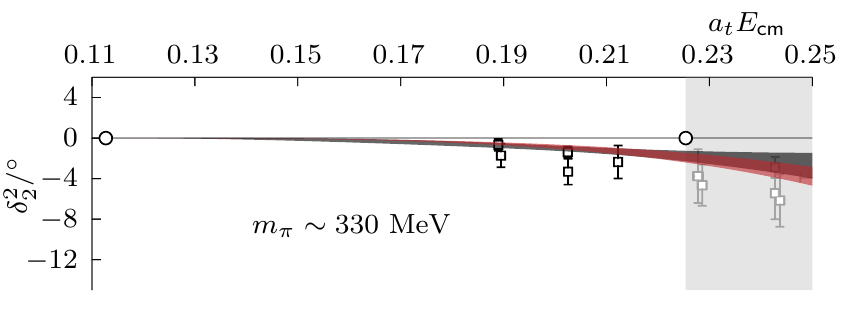}
  \includegraphics[trim=0 10 0 3, width=\columnwidth, clip]{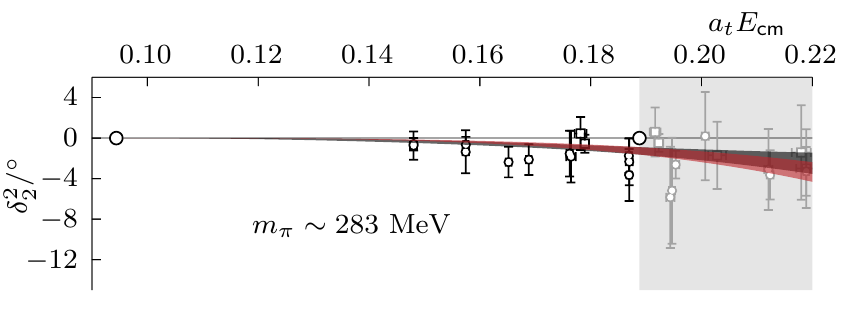}
\caption{
$I=2$ $D$-wave phase-shift for $m_\pi \sim 330$ MeV (top) and $m_\pi \sim 283$ MeV (bottom). Discrete data points come from irreps in which $\ell=2$ is the lowest subduced partial-wave, assuming $\ell\ge 4$ scattering to be negligible. Curves show two illustrative parameterizations: a scattering length form (red) and a conformal mapping with two terms (black), fitted to the full set of energy levels below $\pi\pi\pi\pi$ threshold. The shaded region indicates energies above the $\pi \pi \pi \pi$ threshold. }
\label{fig:I2Dfit}   
\end{figure}
%

%
%
\begin{figure*}
\begin{center}
\resizebox{\textwidth}{!}{
\includegraphics[trim=0 3 0 8,width=\columnwidth]{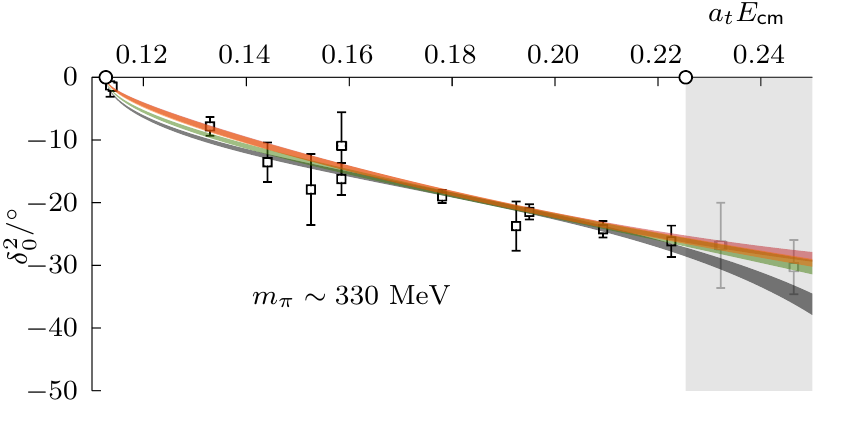}
  \hspace{.1cm}   
  \includegraphics[trim=0 3 0 8,width=\columnwidth]{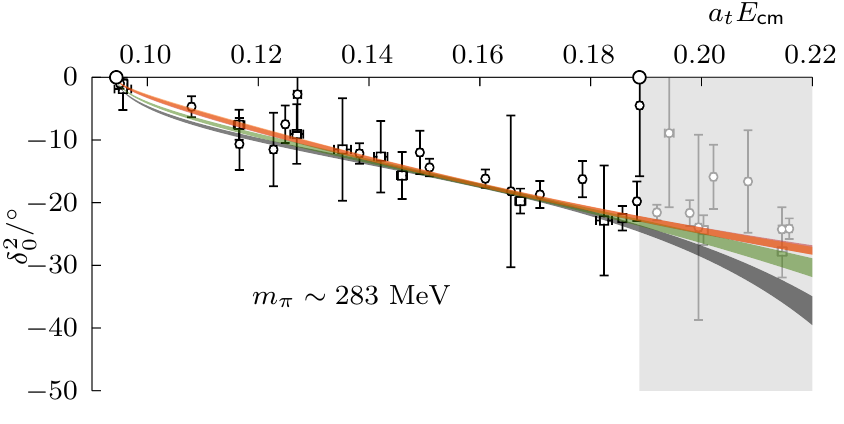}
  }
\resizebox{\textwidth}{!}{
\includegraphics[trim=0 3 0 2,width=\columnwidth]{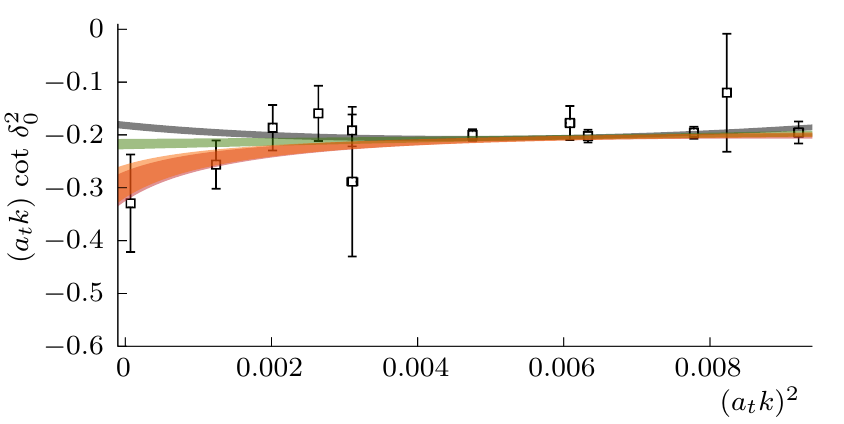}
  \hspace{.1cm} 
\includegraphics[trim=0 3 0 2,width=\columnwidth]{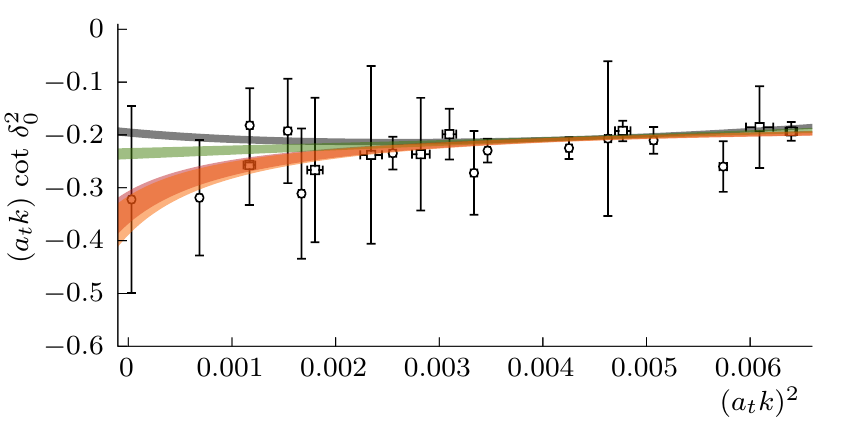}
}
\caption{$I=2$ $S$--wave scattering for $m_\pi\sim330$ MeV (left) and $m_\pi\sim283$ MeV (right). Four example parameterizations are shown: a two-term conformal mapping (black), an effective range expansion with two terms (green), and two choices with an Adler zero fixed at the leading order $\chi$PT location, a two-term conformal mapping (red), and an effective range expansion with two terms (orange).  The enforced presence of the Adler zero can be seen in the deviation of the red and orange curves from nearly flat behavior at threshold in the lower panels. Discrete `data' points with large uncertainties have been removed from the plot for clarity. The shaded region indicates energies above the $\pi \pi \pi \pi$ threshold. Note that, in both cases, the conformal mapping parameterization without an Adler zero produces a relatively poor fit, and for this reason, their scattering lengths are not quoted in~\cref{fig:I2S_SL,fig:I2QMD}.}
\label{fig:I2Sfit}   
\end{center}
\end{figure*}

\subsection{$\pi \pi \to \pi \pi$ $I=2$ scattering}
\label{subsec:I2}

The spectra presented in the previous section can be used to constrain $S$--wave and $D$--wave elastic scattering amplitudes~\footnote{The spectra are compatible with the amplitudes for $\ell \ge 4$ being zero throughout the energy region considered.} using the approach described in~\cref{sec:amplitude-analysis}. Examining energy levels in those irreps whose lowest subduced partial-wave is $\ell=2$, we observe extremely small energy shifts from the non-interacting curves that suggest a very weak interaction. 

Descriptions in terms of parameterizations featuring only a single free parameter, such as a scattering length, lead to good descriptions of the spectra, and as can be seen in~\cref{fig:I2Dfit}, clearly describe a very weak $D$--wave interaction. Adding further parameter freedom does not lead to an improved description of the spectra.

The spectra shown in the lower row of~\cref{fig:I2} are for those irreps in which the $S$--wave is present. These are included together with the spectra in the top row in a $\chi^2$ to obtain descriptions of the $S$-- and $D$--wave amplitudes simultaneously. The $S$--wave amplitudes for several sample parameterization choices are shown in~\cref{fig:I2Sfit}. The principal difference between these various descriptions of the finite-volume spectrum, most of which have $\chi^2/N_\mathrm{dof}\sim 1$, can be observed to be at threshold where the slope of the phase-shift curve, and hence the scattering length, appears to be poorly constrained. This is more clearly seen in the plots of $k \cot \delta^2_0$, where for both pion masses a spread of behaviors at threshold, well outside the statistical uncertainty, is observed. The behaviors fall into two broad categories -- amplitudes where $k \cot \delta^2_0$ is fairly flat at threshold correspond to those which have not been engineered to have an Adler zero below threshold, unlike those which fall at threshold, where an Adler zero was included at the tree-level $\chi$PT location, $s_A = 2 m_\pi^2$.

\begin{figure}[htbp]
\begin{center}
  \includegraphics[width=\columnwidth]{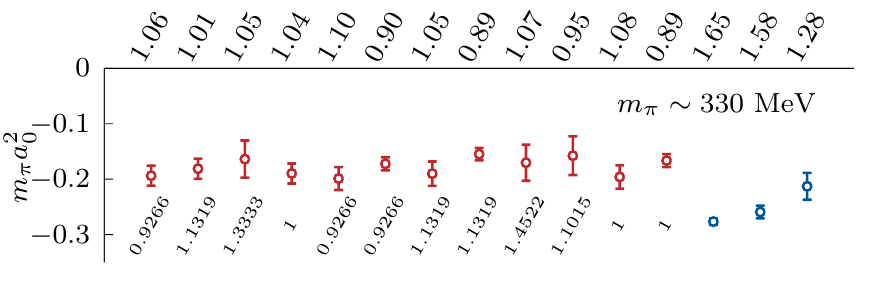}
  \includegraphics[width=\columnwidth]{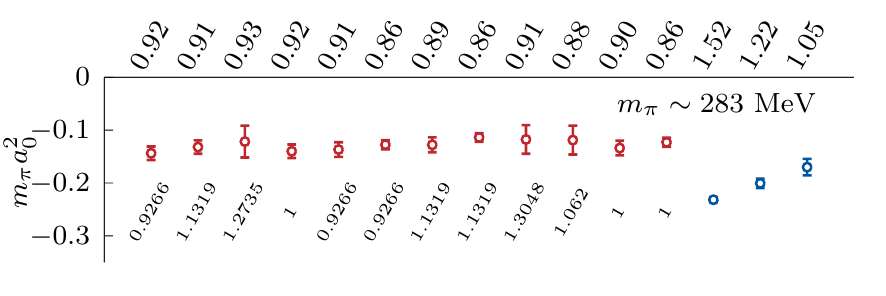}
\caption{ Extracted scattering length for a range of $I=2$ $S$--wave amplitude parameterizations for $m_\pi\sim330 \, \mathrm{MeV}$ (top) and ${m_\pi\sim283\, \mathrm{MeV}}$ (bottom). Each amplitude is labelled by the $\chi^2/N_\mathrm{dof}$, on the top axis, with which it describes the finite-volume spectrum. Red points correspond to amplitudes containing an Adler zero at some location, while blue points lack any enforced subthreshold zero. The numbers below the red points indicate the corresponding Adler zero location, in units of the LO value, $2 m^2_\pi$.}
\label{fig:I2S_SL}   
\end{center}
\end{figure}

Given that the pion masses used in this study are further from the chiral limit than the experimental pion mass, we expect corrections to the tree-level location of an Adler zero that may be significant. As was shown in Ref.~\cite{Garcia-Martin:2011iqs}, dispersive analyses of experimental data suggest that even for the physical pion mass the Adler zero may be displaced from the tree-level location. Motivated by this result, we take the range produced by the ``CFD" dispersive predictions in Ref.~\cite{Garcia-Martin:2011iqs}, and extrapolate it to the pion masses used herein using $s_A = s_A^\mathrm{phys} \left(m_\pi/m_\pi^\mathrm{phys}\right)^2$. We consider descriptions of the finite-volume spectra using amplitudes with Adler zeros fixed at the extremes suggested by this approach, together with some amplitudes for which the Adler zero is allowed to float freely, although these latter choices lead to statistically imprecise results for the amplitude. The location of the enforced Adler zero (or the central value when fitted) for each description is given in~\cref{fig:I2S_SL} as the ratio to the tree-level value, $s_A/2m_\pi^2$.

We plot in~\cref{fig:I2S_SL} the values of $S$-wave scattering length extracted from all parameterizations which provide a reasonable description of the finite-volume spectra, separated between those parameterizations with an Adler zero (of varying location) and those without -- a clear systematic difference is observed, indicating a strong correlation between the location of a subthreshold zero and the value of the scattering length when constrained by only finite-volume energy levels above threshold. This systematic spread is not reduced for the lighter-pion-mass case, despite the fact that we fit over twice the data points than for the heavier mass.

\begin{figure}[htbp]
\begin{center}
\includegraphics[width=\columnwidth]{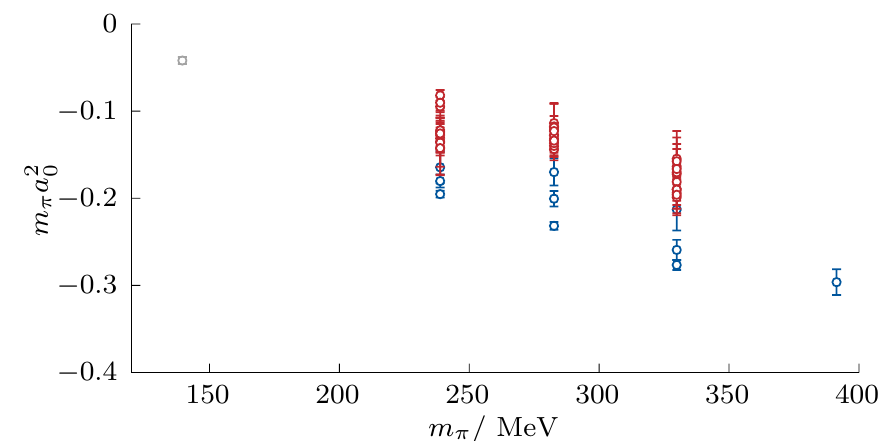}
\caption{$I=2$ $S$--wave scattering length extracted from parameterizations describing finite-volume spectra at four pion masses. Red points indicate amplitudes that feature an Adler zero, while blue points lack an enforced subthreshold zero. The result of dispersive analysis applied to experimental data~\cite{Garcia-Martin:2011iqs} is shown by the gray point.
}
\label{fig:I2QMD}   
\end{center}
\end{figure}

In~\cref{fig:I2QMD} we show the pion mass evolution of the $I=2$ $S$-wave scattering length across four pion masses computed with the same lattice action. The rightmost point, taken from Ref.~\cite{Dudek:2012gj} reflects an average over several parameterizations in which Adler zeroes were not enforced. The leftmost points, at $m_\pi \sim 239$ MeV, show an analysis of the same type followed by this paper that has not been previously published~\footnote{Details of this analysis are provided in~\cite{Rodas:2023twk}.}. It is clear that the slope of the variation with pion mass is very sensitive to the existence, or not, of an Adler zero. We return in this paper's conclusions to whether the presence and exact location of the Adler zero, which lies far from the region of constraint provided by finite-volume energy levels, can be resolved using only lattice QCD data.

\pagebreak
\vspace*{1cm}
\begin{figure*}[htbp]
\begin{center}
\resizebox{\textwidth}{!}{
  \includegraphics{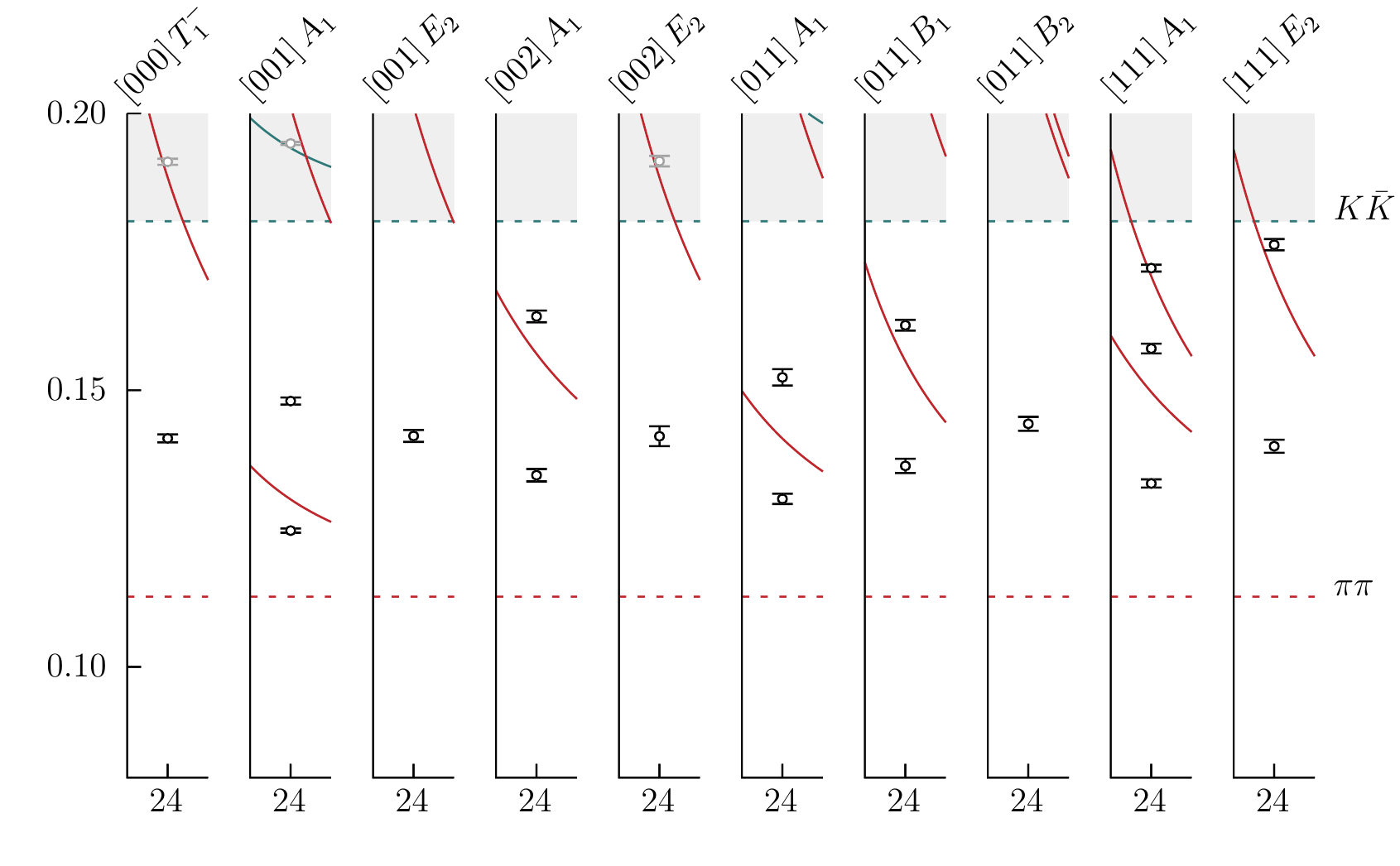}
  \includegraphics{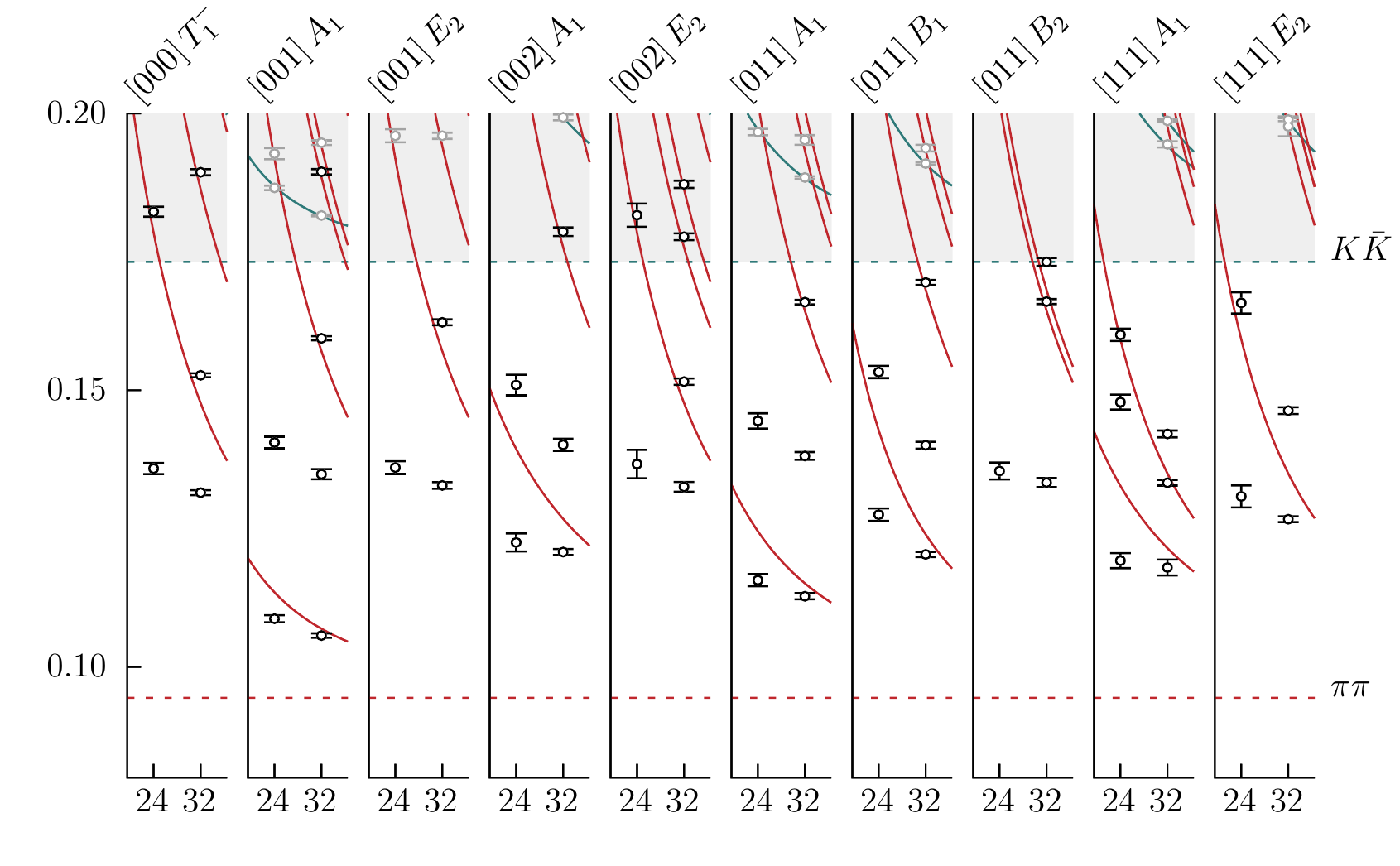}
}
\caption{
$I=1$ finite-volume spectra, $a_t E_\mathsf{cm}$, by irrep, against $L/a_s$, for $m_\pi \sim 330$ MeV (left) and $m_\pi \sim 283$ MeV (right). Red/green curves show $\pi\pi$/$K\bar{K}$ non-interacting energies. Levels appearing on top of $K\bar K$ non-interacting energies are not considered in our elastic fits to data. }

\label{fig:I1spec}   
\end{center}
\end{figure*}

\section{$\pi \pi \to \pi \pi$ $I=1$}
 \label{sec:I1pipi}

The $I=1$ channel contains the $P$--wave $\rho$ resonance which appears below the $K\bar{K}$ threshold, while the \mbox{$F$--wave} amplitude is expected to be featureless and very weak across the elastic region. In order to reliably determine the finite-volume spectrum up to slightly above the $K\bar{K}$ threshold, we make use of a large basis of single hadron operators, $\pi\pi$--like operators, and $K\bar{K}$--like operators.

\subsection{$I=1$ finite-volume spectra}
\label{subsec:I1FV}

Figure~\ref{fig:I1spec} shows the extracted spectra for the two pion masses considered in this calculation, where large departures from the $\pi\pi$ non-interacting energies (red curves) are observed, indicative of strong interactions. The isolated `extra' levels near $a_t E_\mathsf{cm} \sim 0.14$ suggest a narrow resonance in that energy region. At higher energies, the extracted finite-volume spectra lie very close to the non-interacting energies (including those corresponding to $K\bar{K}$) suggesting that the scattering amplitude may be featureless above the resonance.

In total, we extracted 23 levels for $m_\pi \sim 330$ MeV and $95$ levels for $m_\pi \sim 283$ MeV, of which 17 and 42 are below the $K \bar K$ threshold, respectively. Examination of the operator overlaps for states above the $K\bar{K}$ threshold suggests that there appears to be no significant coupling between the $\pi\pi$ and $K\bar{K}$ channels, indicating that an analysis of elastic scattering above threshold, retaining only those levels with overlap onto $\pi\pi$ operators, may be successful. This appears to be essentially the same situation as was observed for $m_\pi \sim 239$ MeV in Ref.~\cite{Wilson:2015dqa}~\footnote{Referred to in that paper as $m_\pi \sim 236$ MeV. An improved extraction of the pion mass and the $\Omega$ baryon mass used to set the scale provide the newer value.}, where coupled channel analysis showed negligible $\pi\pi, K\bar{K}$ channel coupling over a significant energy region above threshold.

\subsection{$\pi \pi \to \pi \pi$ $I=1$ scattering}
\label{subsec:I1}

\begin{figure}[htbp]
\begin{center}
  \includegraphics[width=\columnwidth]{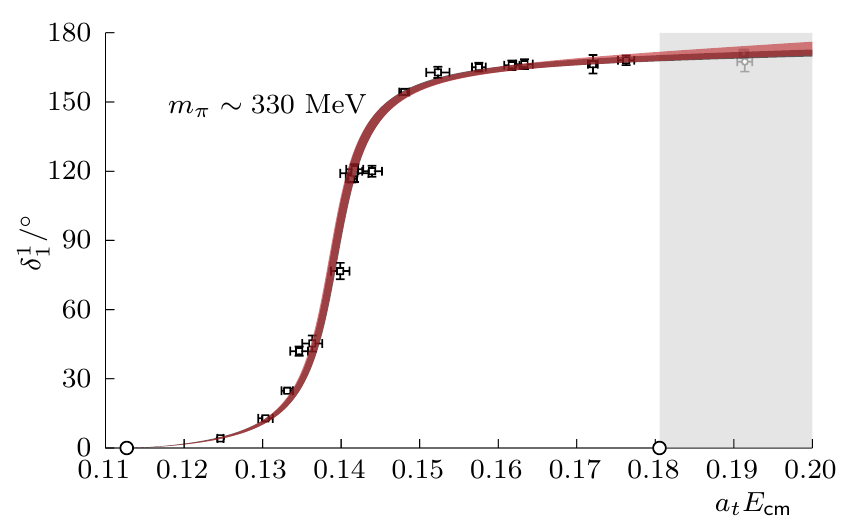}
  
  \includegraphics[width=\columnwidth]{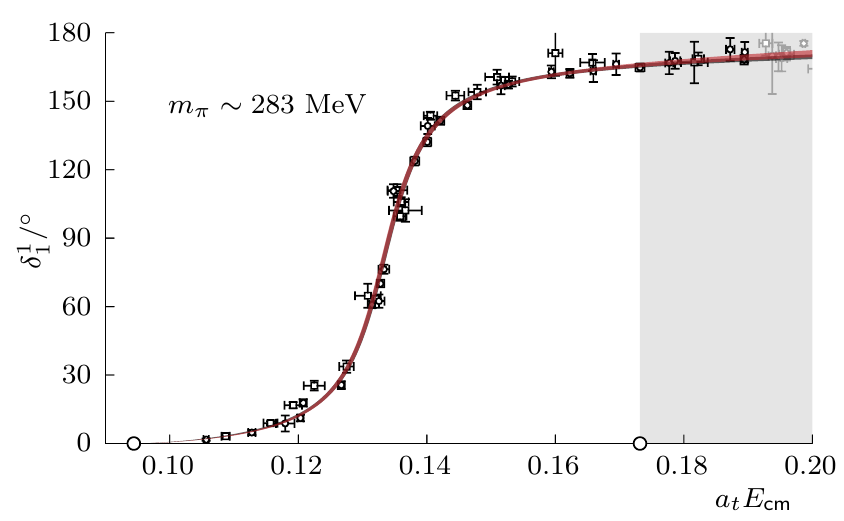}
\caption{
$I=1$ $P$--wave phase-shift for $m_\pi \sim 330$ MeV (top) and $m_\pi \sim 283$ MeV (bottom). A parameterization using a conformal mapping with a resonance enforcing $F^I_\ell(s)$ factor shown by the black curve, and a $K$--matrix with a single pole plus a constant shown by the red curve. Discrete `data' points with large uncertainties have been removed from the plot for clarity. The shaded region indicates  energies above the $K \bar K$ threshold.
}
\label{fig:I1Pfit}   
\end{center}
\end{figure}

As explained above, we restrict ourselves to an elastic analysis in this manuscript, and the extracted spectra indicate that the $F$-wave amplitude is negligible relative to the $P$-wave in the region of interest. The $\ell=3$ angular momentum barrier factor suppresses the low-energy interactions, and the only resonance with those quantum numbers that decays to $\pi \pi$ is a $\rho_3$, which is expected to appear far above the energy region we consider. Thus, in this case, each energy level can be used to determine a discrete value of $\delta^1_1(s)$, as plotted in~\cref{fig:I1Pfit}. The behavior for each pion mass is clearly that of a narrow resonance, and we consider elastic amplitude parameterizations which describe the finite-volume spectra up to $a_t E_\mathsf{cm} = 0.19$.

A Breit-Wigner form,~\cref{eq:bw}, is found to describe the finite-volume spectra reasonably, with parameters
\begin{center}
	\begin{tabular}{rll}
	$m_\mathrm{BW} =$                         & $0.13978\, (51) \cdot a_t^{-1}$   &
	\multirow{2}{*}{ $\begin{bmatrix*}[r] 1 &  0.08  \\
					                    & 1    \end{bmatrix*}$ } \\
			$g_\mathrm{BW} = $                  & $5.664\,(104)$   & \\[1.3ex]
			\multicolumn{2}{r}{$\chi^2/N_{\text{dof}}=\frac{14.42}{17-2}=0.96$.}
		\end{tabular}
	\end{center}
	\vspace{-1.0cm}
	\begin{equation}\label{eq:bw_850}\end{equation}
for $m_\pi \sim 330 \,\mathrm{MeV}$, and
\begin{center}
	\begin{tabular}{rll}
	$m_\mathrm{BW} =$                         & $0.13440\, (34) \cdot a_t^{-1}$   &
	\multirow{2}{*}{ $\begin{bmatrix*}[r] 1 & 0.00  \\
					                    & 1    \end{bmatrix*}$ } \\
			$g_\mathrm{BW} = $                  & $5.560\,(61)$   & \\[1.3ex]
			\multicolumn{2}{r}{$\chi^2/N_{\text{dof}}=\frac{41.34}{49-2}=0.88$.}
		\end{tabular}
	\end{center}
	\vspace{-1.0cm}
	\begin{equation}\label{eq:bw_856}\end{equation}
for $m_\pi \sim 283 \,\mathrm{MeV}$. The matrices illustrate the statistical correlation between the parameters. In both cases, the Breit-Wigner mass and coupling parameters are essentially uncorrelated.

Considering a wider variety of amplitude parameterizations, including $K$--matrix forms and conformal expansions, we can find examples that describe the data with slightly improved $\chi^2/N_\mathrm{dof}$, but all successful descriptions show compatible phase-shift energy dependence in the region of the resonance. In the next section, we will examine the variation of the $\rho$ resonance pole with parameterization choice.

\begin{figure}[htbp]
\begin{center}
  \includegraphics[width=\columnwidth]{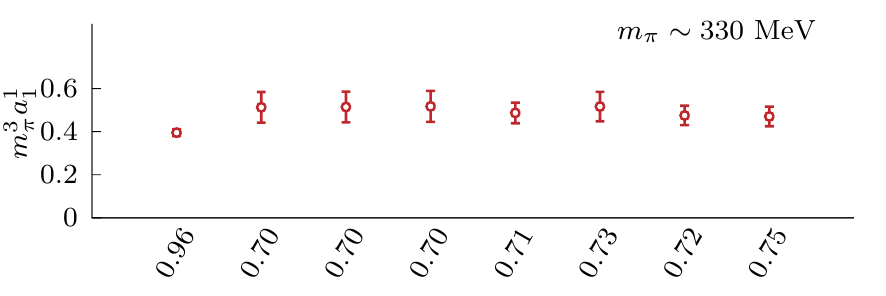}
  
  \includegraphics[width=\columnwidth]{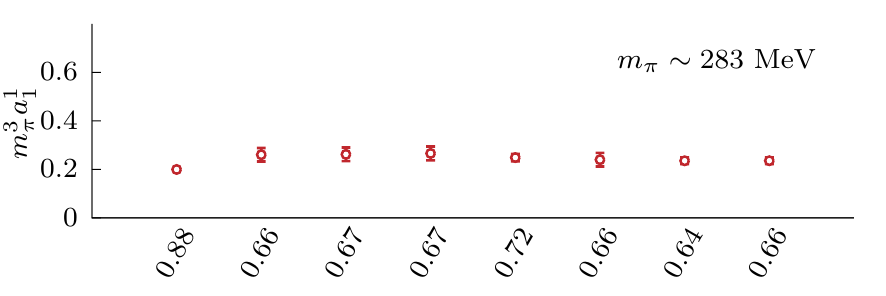}
\caption{
Extracted scattering length for a range of $I=1$ $P$--wave amplitude parameterizations for ${m_\pi\sim330 \,\mathrm{MeV}}$ (top) and ${m_\pi\sim283 \,\mathrm{MeV}}$ (bottom). Each amplitude is labeled by the $\chi^2/N_\mathrm{dof}$ with which it describes the finite-volume spectrum. 
}
\label{fig:I1S_SL}   
\end{center}
\end{figure}

The amplitude at threshold is characterized by a scattering `length', defined via $k^3 \cot \delta^1_1 =  \frac{1}{a^1_1}$, and as seen in~\cref{fig:I1S_SL}, amplitudes capable of describing the finite-volume spectrum with the smallest $\chi^2$ values have compatible values of this parameter. The first entry plotted for each pion mass corresponds to the Breit-Wigner fit, and such a form is not expected to provide a faithful description of amplitudes away from the resonance that is being parameterized, and hence this form may not describe the threshold behavior accurately.

\subsection{The $\rho$ resonance}
 \label{subsec:rho}

\begin{figure*}[htbp]
\begin{center}
\resizebox{\textwidth}{!}{
  \includegraphics{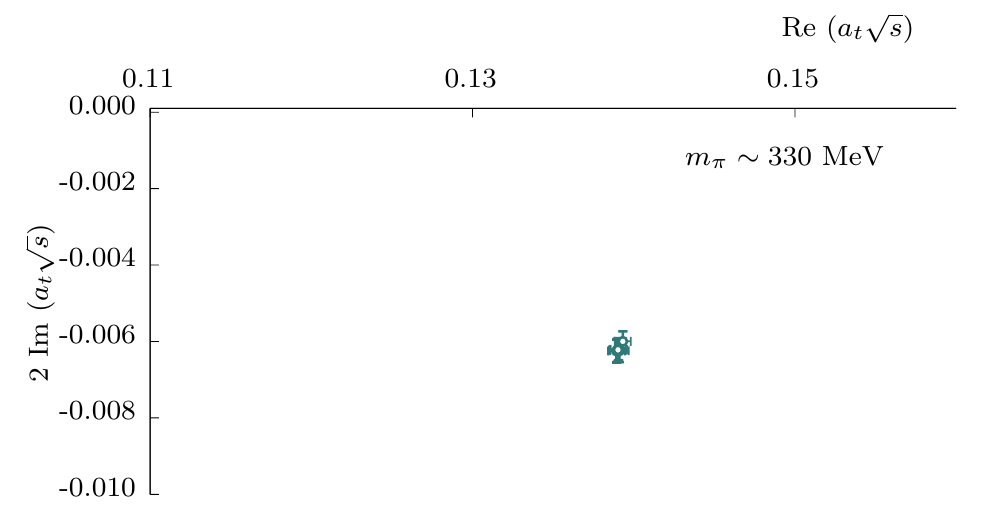}
  \hspace{.1cm} 
  \includegraphics{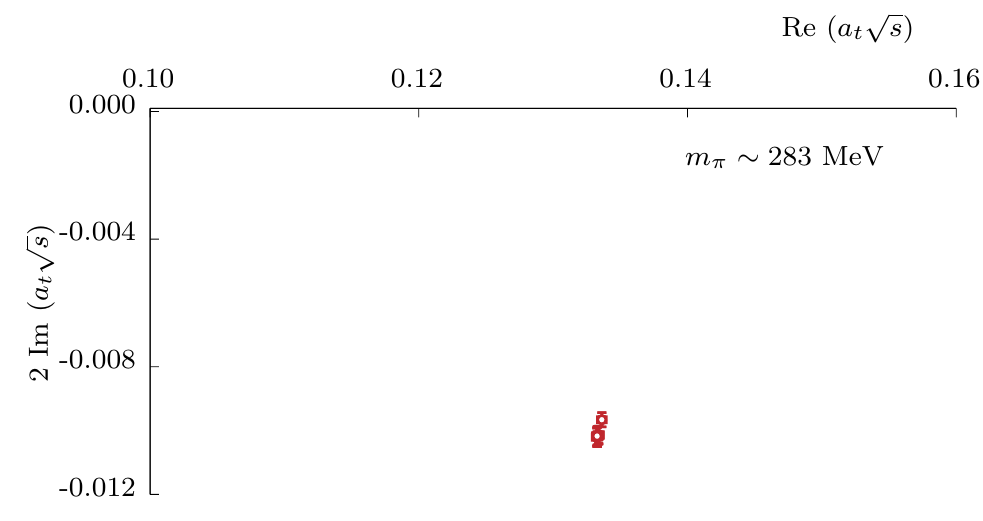}
}
\caption{
Extracted $\rho$ resonance pole location for each $I=1$ $P$-wave parameterization found capable of describing the finite-volume spectra for $m_\pi\sim330$ MeV (left) and $m_\pi\sim283$ MeV (right).
}
\label{fig:I1Ppoles}   
\end{center}
\end{figure*}

In the case of the $I=1$ $P$--wave, for both pion masses, a pole singularity lying near the real axis is found on the second Riemann sheet for every parameterization that successfully describes the finite-volume spectra. The pole location for each parameterization is plotted in~\cref{fig:I1Ppoles} where we observe very little scatter, indicating that the lattice spectra precisely determine the mass and width of the $\rho$ resonance at these pion masses without significant amplitude parameterization dependence.

\begin{figure*}[htbp]
\begin{center}
\resizebox{\textwidth}{!}{
  \includegraphics{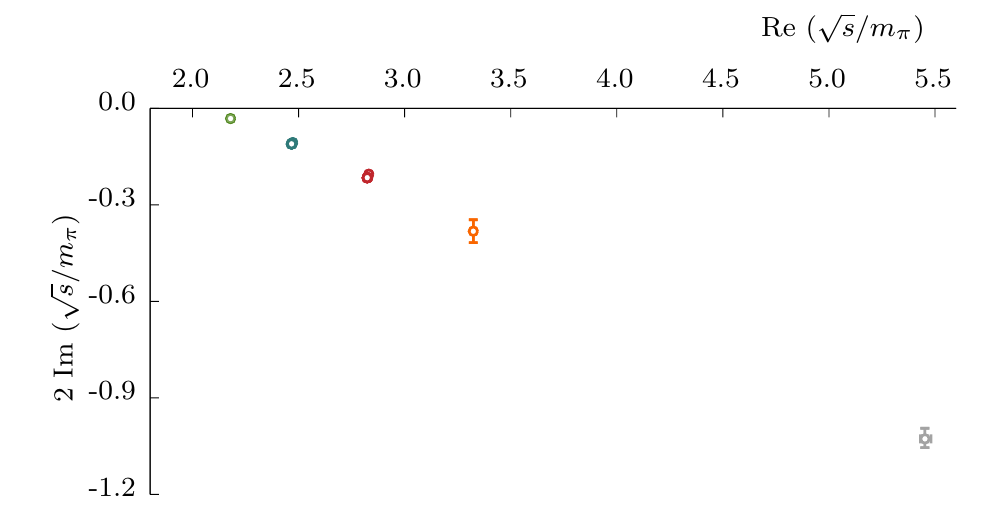}
  \hspace{.1cm} 
  \includegraphics{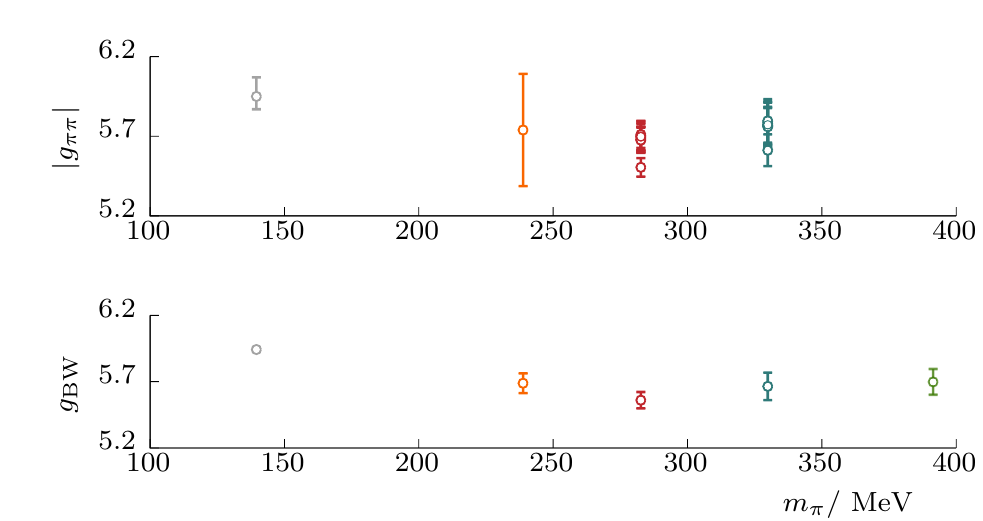}
}
\caption{ Left: $\rho$ resonance pole location with varying pion mass from this calculation (blue and red points) and from calculations on lattices with the same action~\cite{Dudek:2012xn, Wilson:2015dqa} (green, orange).  Right: Magnitude of the complex $\rho$ resonance pole coupling, as defined in~\cref{gNorm}, and the real coupling, $g_\mathrm{BW}$, extracted when a Breit-Wigner, Eqn~\ref{eq:bw}, is used to describe the spectrum. The uncertainties on the pole location and pole couplings quoted from Ref.~\cite{Wilson:2015dqa} (orange points) are a rather conservative average over a large number of parameterizations, including several which include the $K\bar{K}$ coupled-channel region. For pole properties, the ``Roy" result of dispersive analysis of experimental data~\cite{Garcia-Martin:2011nna} is shown in gray, while the physical value of $g_\mathrm{BW}$ comes from the neutral $e^+ e^-$ mode listed in the PDG~\cite{ParticleDataGroup:2022pth}.
}
\label{fig:rhopoles}   
\end{center}
\end{figure*}

These $\rho$ pole results supplement those obtained in Refs.~\cite{Dudek:2012xn,Wilson:2015dqa} at $m_\pi \sim 391, 239$ MeV, and in~\cref{fig:rhopoles} we present the evolution of the pole position and pole residue coupling (defined in~\cref{gNorm}) with varying pion mass. As expected the $\rho$ becomes heavier as the light quark mass increases and narrower as the phase-space for decay to two pions decreases. The coupling appears to be consistent with being constant across the range of pion masses considered. These results agree with the expectations for an `ordinary $q \bar q$ meson' as defined in Ref.~\cite{Pelaez:2006nj}, and agree with predictions made for the quark mass trajectory of the $\rho$  in unitarized chiral perturbation theory models~\cite{Hanhart:2008mx,Pelaez:2010fj,Nebreda:2010wv,Chen:2012rp,Hu:2017wli,Molina:2020qpw,Niehus:2020gmf}.

\pagebreak
\section{$\pi \pi \to \pi \pi$ $I=0$}
 \label{sec:I0pipi}

Isospin--0 $\pi\pi$ scattering in $S$--wave is not simple to describe, being neither weak nor dominated by a single narrow resonance. At the physical pion mass, despite there being relevant scattering data available for over forty years, it is only recently that the role of the broad $\sigma$ resonance has been confirmed with certainty~\cite{Caprini:2005zr,Garcia-Martin:2011nna,Moussallam:2011zg}. Recent consideration of this scattering channel using first-principles lattice QCD showed a clear $\sigma$ \emph{bound-state} when $m_\pi \sim 391$ MeV, and evidence for a broad $\sigma$ resonance (albeit with significant parameterization dependence) when $m_\pi \sim 239$ MeV~\cite{Briceno:2016mjc}. We will observe in the current calculation that between the two pion masses considered here the $\sigma$ undergoes a dramatic change in form.

\begin{figure*}[htbp]
\begin{center}
\resizebox{\textwidth}{!}{
  \includegraphics{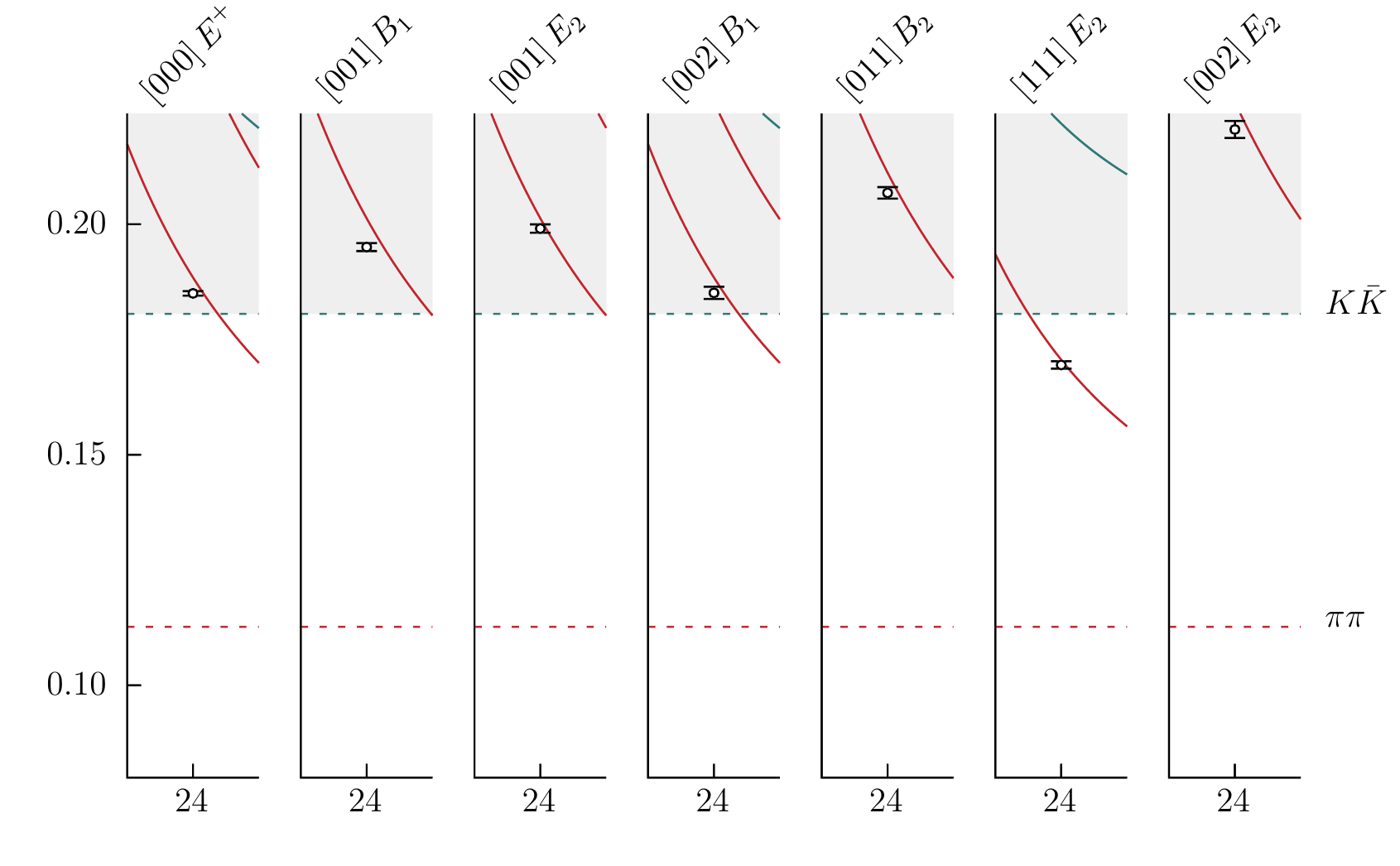}
  \includegraphics{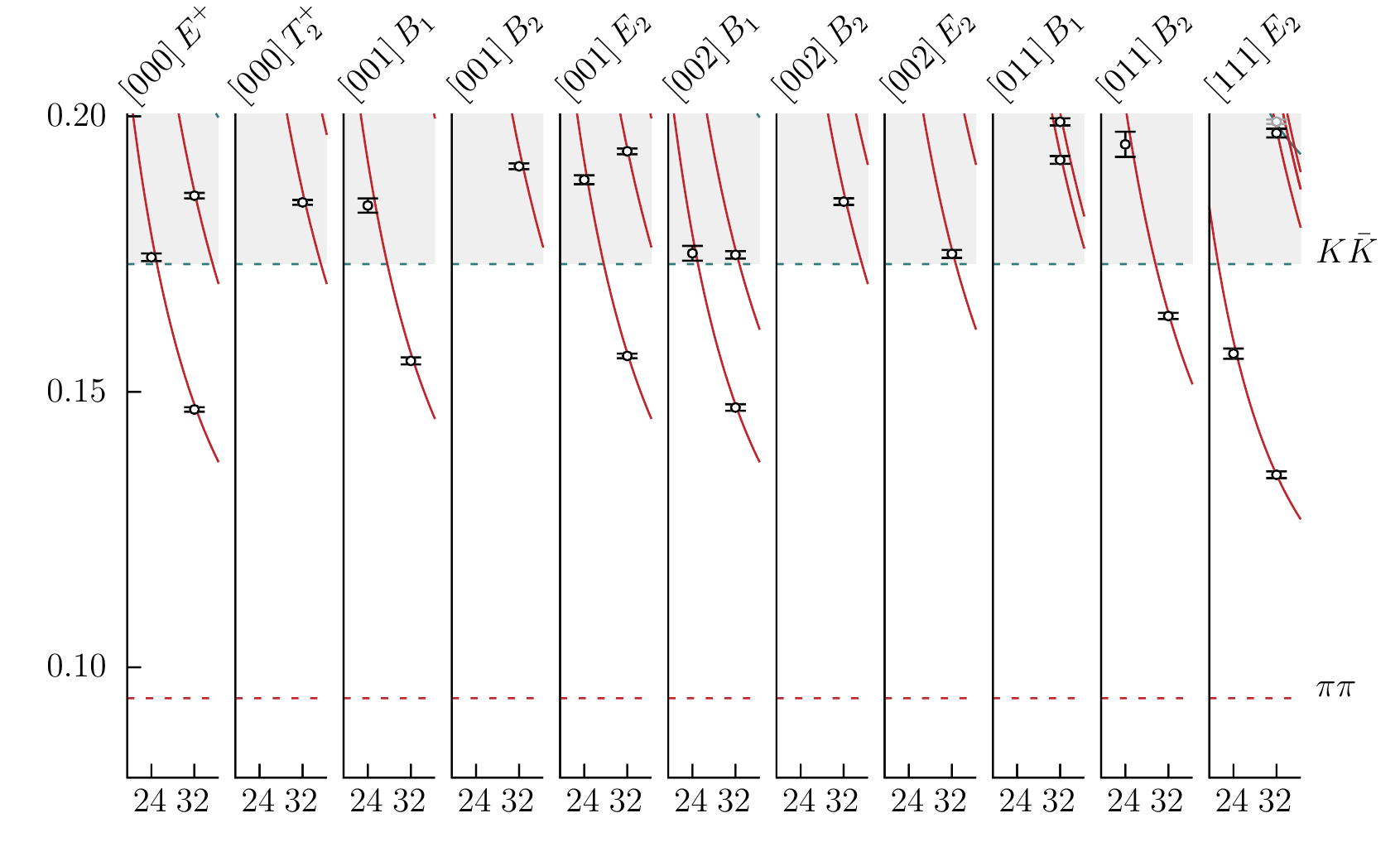}
}
\resizebox{\textwidth}{!}{
  \includegraphics{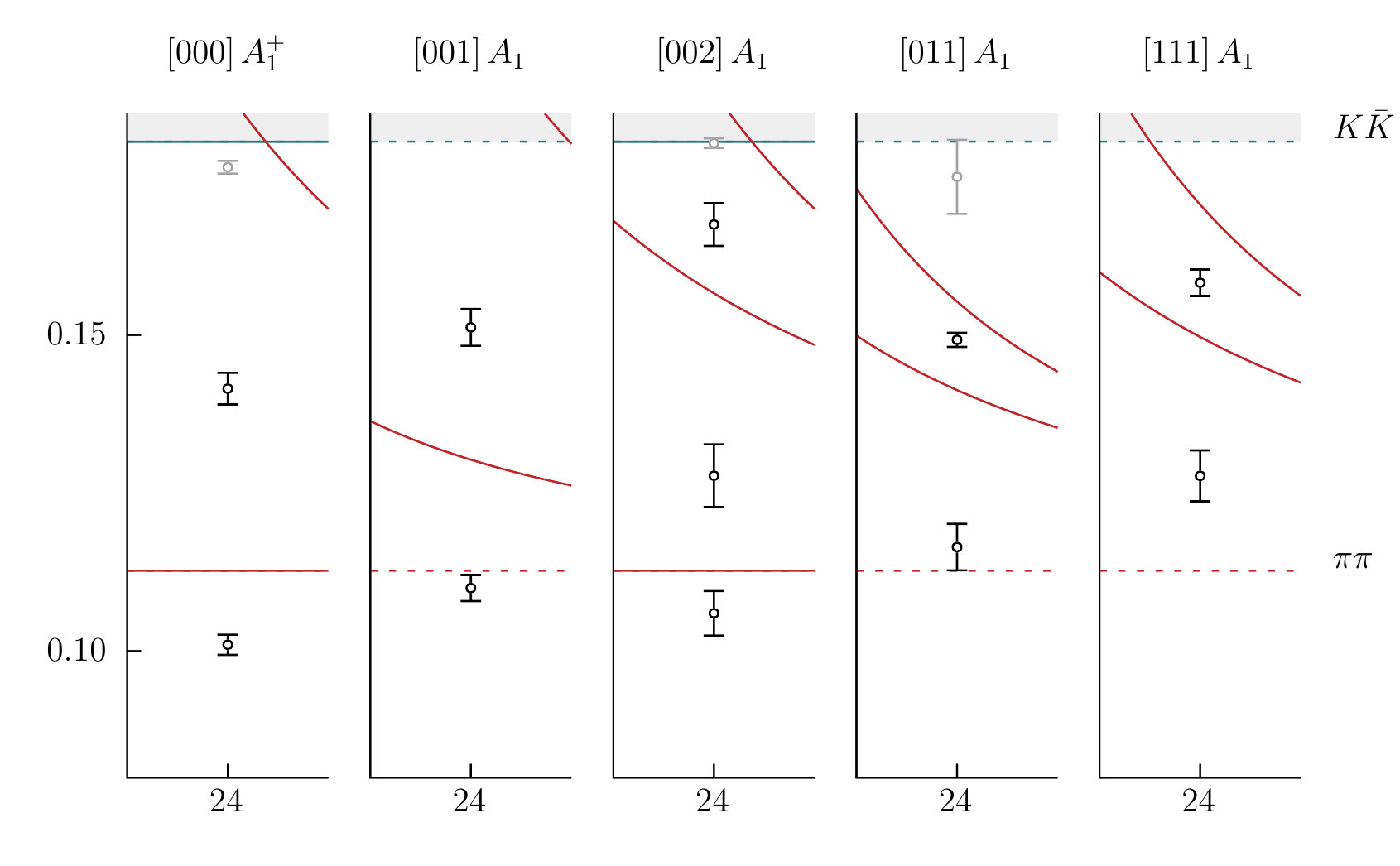}
  \includegraphics{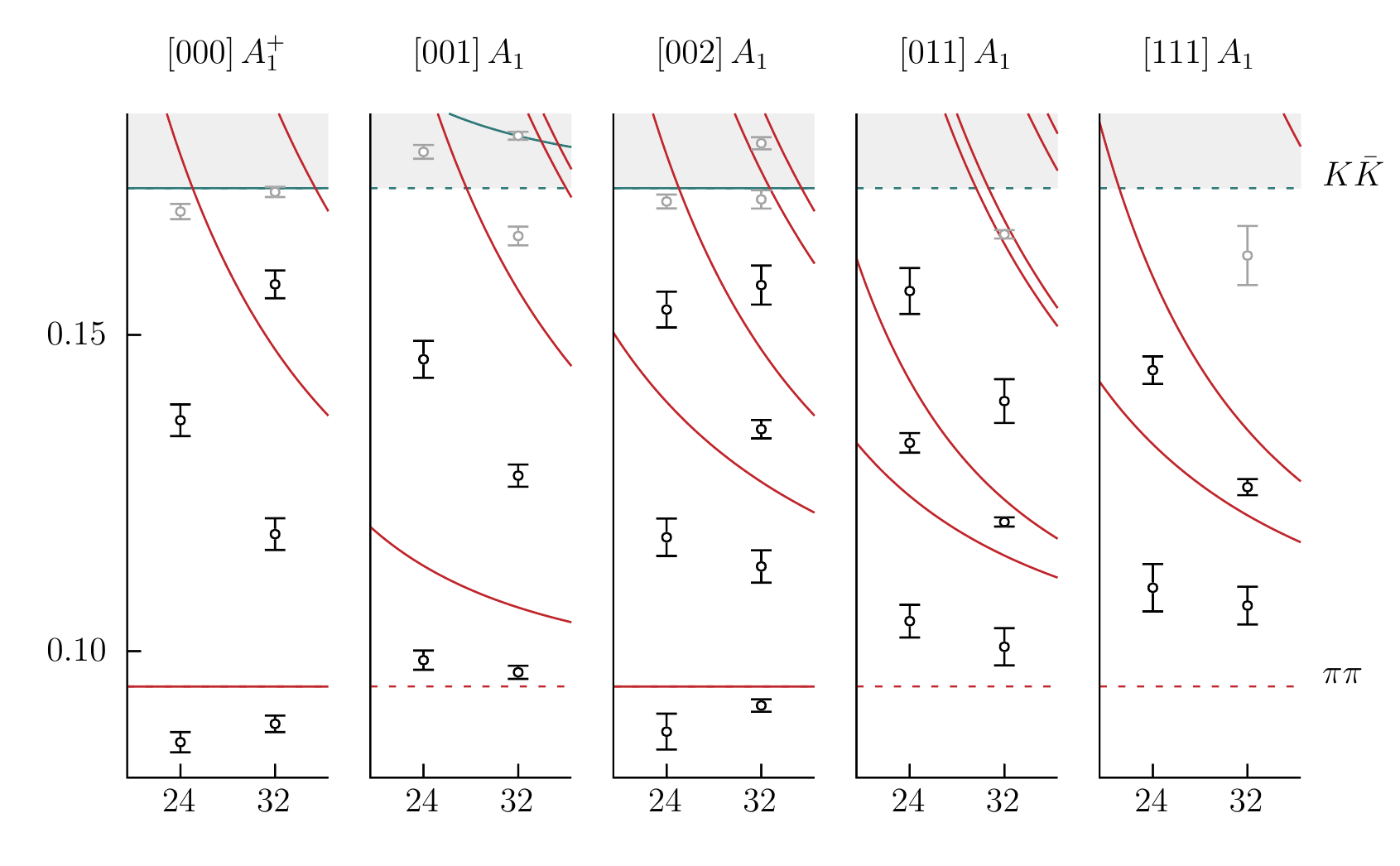}
}
\caption{
$I=0$ finite-volume spectra, $a_t E_\mathsf{cm}$, by irrep, against $L/a_s$, for $m_\pi \sim 330$ MeV (left) and $m_\pi \sim 283$ MeV (right). Upper panel irreps have $D$-wave as their leading partial-wave, while those in the lower panel have $S$-wave leading. Red/green curves show $\pi\pi$/$K\bar{K}$ non-interacting energies.
}
\label{fig:I0}   
\end{center}
\end{figure*}

\subsection{$I=0$ finite-volume spectra}
\label{subsec:I0FV}

Spectra were extracted from correlator matrices computed using a basis of single-hadron operators, $\pi\pi$-like operators, $K\bar{K}$-like operators, and some $\eta\eta$-like operators (for the lighter pion mass, larger volume lattice). The energies are shown in~\cref{fig:I0}, where it is clear that there are large departures from the non-interacting $\pi\pi$ energies in those irreps containing subduction of the $\pi\pi$ $S$--wave suggesting strong scattering, while those irreps having $D$--wave as their leading partial-wave show only small downward shifts indicative of mild attraction. \mbox{$D$--wave} resonances, $f_2, f_2'$, are expected to lie at significantly higher energy, well into the coupled-channel region~\cite{Briceno:2017qmb}.

Even though the $I=0$ correlation functions receive vital contributions from relatively noisy diagrams featuring complete quark-line annihilation, the use of distillation leads to high-quality signals, and the extracted energy levels are of high statistical precision. The error bars on the points plotted in~\cref{fig:I0} also include systematic errors originating from fitting variations, which are modest in most cases.

\begin{figure}[htbp]
\begin{center}
  \includegraphics[width=\columnwidth]{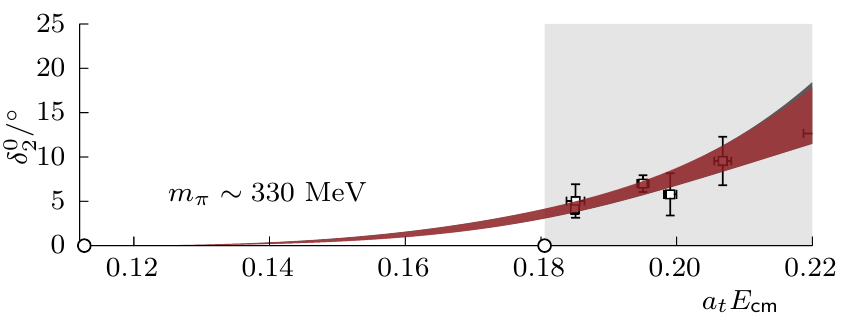}
  \includegraphics[width=\columnwidth]{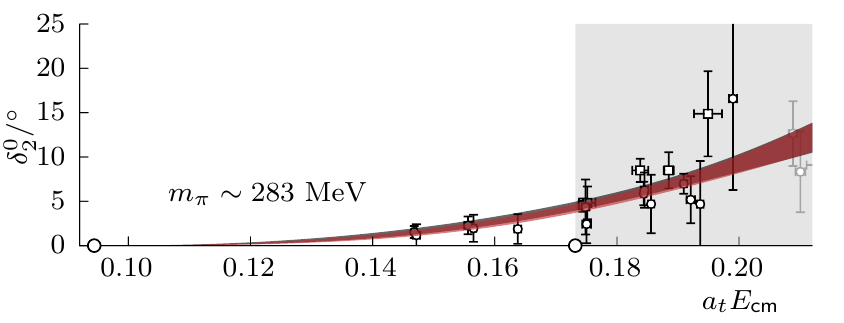}
\caption{
$I=0$ $D$-wave phase-shift for $m_\pi \sim 330$ MeV (top) and $m_\pi \sim 283$ MeV (bottom). Discrete data points come from irreps in which $\ell=2$ is the lowest subduced partial-wave, assuming $\ell\ge 4$ scattering to be negligible. 
Curves show two illustrative parameterizations: an effective range expansion with two terms (red) and a conformal mapping with two terms (black). The shaded region indicates energies above the $K \bar K$ threshold.
}
\label{fig:I0Dfit}   
\end{center}
\end{figure}

In total, there are 23 levels for $m_\pi \sim 330$ MeV and $75$ levels for $m_\pi \sim 283$ MeV. For the $D$--wave dominated irreps, there is no evidence of coupling between $\pi\pi$ and $K\bar{K}$, and a description in terms of purely elastic $\pi\pi$ scattering, even above the $K\bar{K}$ threshold will prove to be successful. The $S$--wave dominated irreps on the other hand cannot be described so simply, and we consider only energy levels lying some way below the $K\bar{K}$ threshold, where channel coupling is expected to turn on rapidly. For $m_\pi \sim 330$ MeV we use 18 energy levels, and 48 for $m_\pi \sim 283$ MeV.

\subsection{$\pi \pi \to \pi \pi$ $I=0$ scattering}
\label{subsec:I0}

There is no evidence in the computed spectra that amplitudes with $\ell > 2$ are required over the energy region we are considering, and as seen in~\cref{fig:I0Dfit}, even the $D$-wave amplitude is only very mildly attractive.

\begin{figure*}[htbp]
\begin{center}
\resizebox{\textwidth}{!}{
  \includegraphics[trim=0 3 0 8]{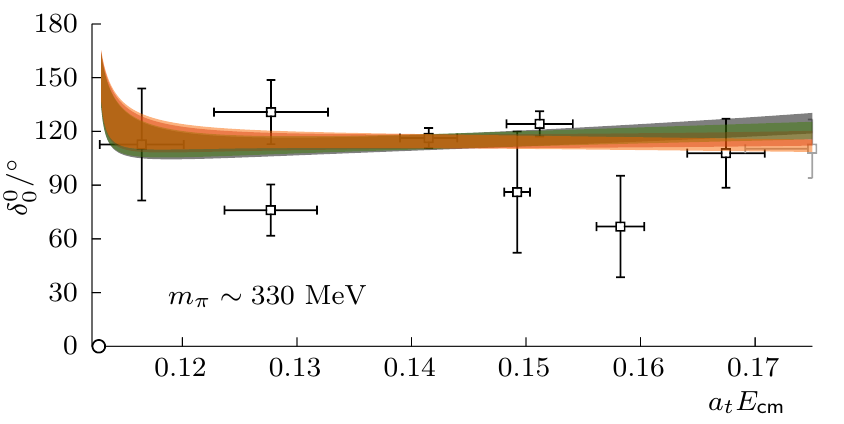}
  \hspace{.1cm} 
  \includegraphics[trim=0 3 0 8]{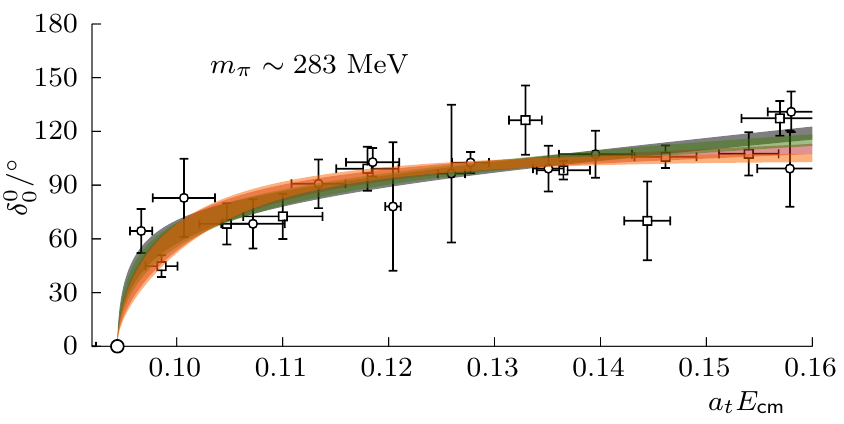}
}
\resizebox{\textwidth}{!}{
  \includegraphics[trim=0 5 0 2]{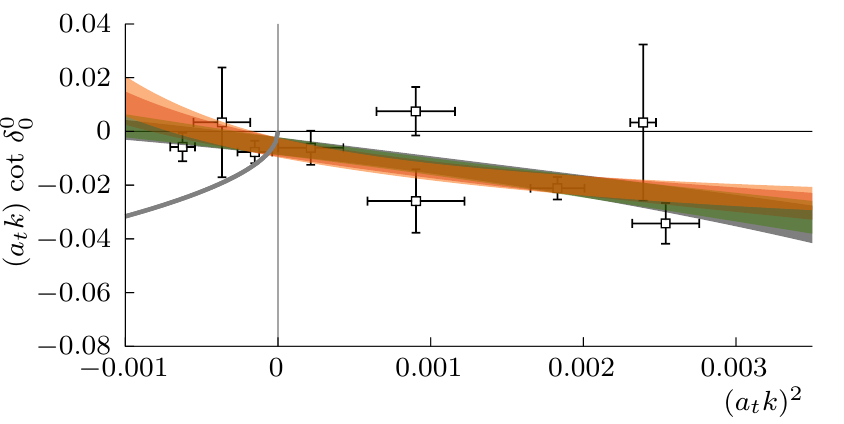}
  \hspace{.1cm} 
  \includegraphics[trim=0 5 0 2]{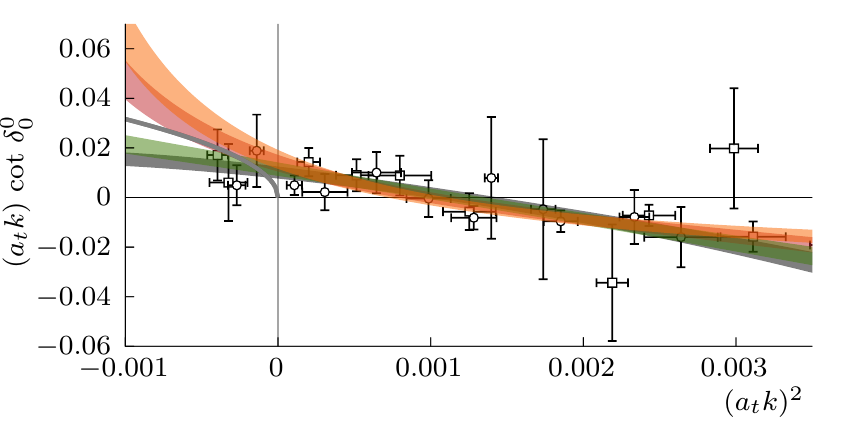}
}
\caption{
$I=0$ $S$--wave scattering for $m_\pi\sim330$ MeV (left) and $m_\pi\sim283$ MeV (right). Four example parameterizations are shown: a two-term conformal mapping (black), an effective range expansion with two terms (green), and two choices with an Adler zero fixed at the leading order $\chi$PT location, a two-term conformal mapping (red), and an effective range expansion with two terms (orange). In the bottom panels, the gray curves show $\mp \sqrt{-k^2}$ below threshold -- intersection of the $k \cot \delta^0_0$ curves with these indicate the location of a bound-state or a virtual bound-state, respectively.}
\label{fig:I0Sfit}   
\end{center}
\end{figure*}

The $S$--wave amplitudes, shown in~\cref{fig:I0Sfit}, provide our first example of amplitudes whose description is not obvious, and where the behavior changes dramatically between the two pion masses considered. The lighter pion mass shows a phase-shift increasing with a moderate slope from threshold, and when plotted as $k \cot \delta^0_0$, a crossing of threshold at a small positive value, indicating a large positive scattering length. The heavier pion mass shows a qualitatively different energy dependence, having an approximately flat phase-shift above threshold, and a $k \cot \delta^0_0$ threshold crossing at a small negative value, indicating a large negative scattering length.

\begin{figure}[htbp]
\begin{center}
\includegraphics[trim=0 2 0 12,width=\columnwidth]{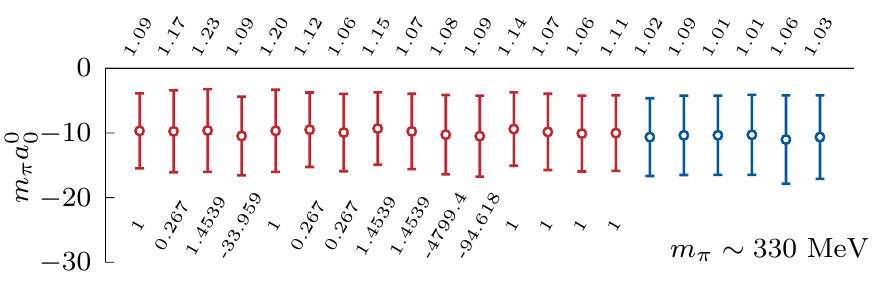} 
\includegraphics[trim=0 3 0 0,width=\columnwidth]{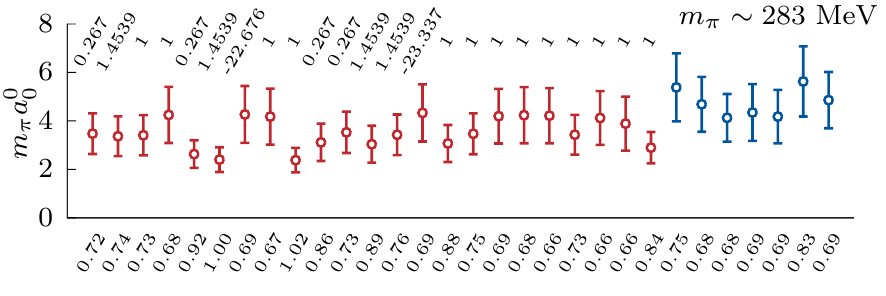}
\caption{Extracted scattering length for a range of $I=0$ $S$--wave amplitude parameterizations for $m_\pi\sim330 \, \mathrm{MeV}$ (top) and ${m_\pi\sim283\, \mathrm{MeV}}$ (bottom). Each amplitude is labeled by the $\chi^2/N_\mathrm{dof}$, on the top(bottom) axis, with which it describes the finite-volume spectrum. Red points correspond to amplitudes containing an Adler zero at some location, while blue points lack any enforced subthreshold zero. The numbers closest to the red points indicate the corresponding Adler zero location, in units of the LO value, $m_\pi^2/2$. Negative values come from those cases where the zero location is a free parameter, and these values have large uncertainties.}
\label{fig:I0S_SL}   
\end{center}
\end{figure}

The plots of $k \cot \delta^0_0$ show clearly the presence of a systematic variation with choice of parameterization. A wide range of forms of the type described in Section~\ref{sec:amplitude-analysis} is used, including cases with and without an Adler zero. As was the case for $I=2$ $S$--wave analysis, we explore Adler zeroes fixed at the leading order location, $s_A = \tfrac{1}{2}m_\pi^2$, and varying between the dispersive ``CFD" predictions in Ref.~\cite{Garcia-Martin:2011iqs} (appropriately scaled for the changed pion mass), and finally, allowing the zero location to float as a free parameter. The locations of the different Adler zeroes are given in~\cref{fig:I0S_SL}.

The four illustrative cases presented in~\cref{fig:I0Sfit} (right) indicate a slight sensitivity to the presence of an Adler zero, likely reduced relative to the $I=2$ case by virtue of the zero being further below threshold. The energy levels below threshold do not obviously suggest a preference either way for an Adler zero.

\begin{figure}[htbp]
\begin{center}
\resizebox{\columnwidth}{!}{
  \includegraphics{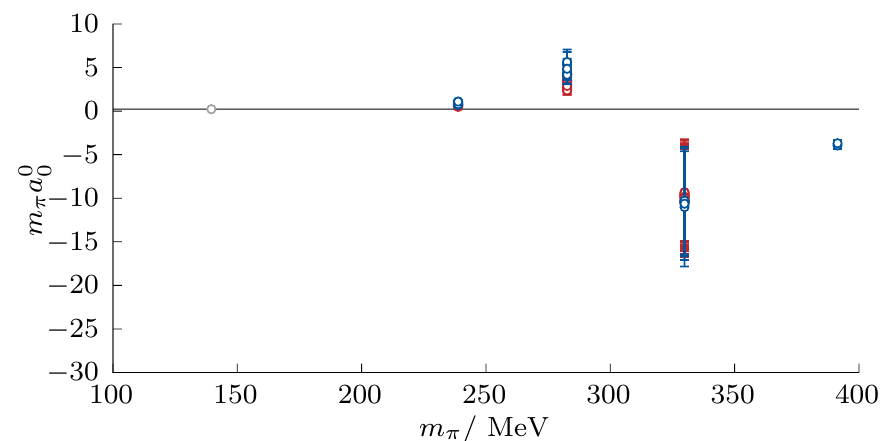}
}
\caption{ $I=0$ $S$--wave scattering length at four pion masses (this paper and Ref.~\cite{Briceno:2016mjc}). Red points correspond to parameterizations featuring an Adler zero, while blue points have no enforced subthreshold zero. 
The result of dispersive analysis applied to experimental data~\cite{Garcia-Martin:2011iqs} is shown by the gray point.
}
\label{fig:I0_SL_quarkmass}   
\end{center}
\end{figure}

Figure~\ref{fig:I0S_SL} shows the scattering length for each parameterization choice that successfully describes the finite-volume spectra. It is clear that at the heavier pion mass, the presence, or not, of an Adler zero has no impact on the value of the scattering length, and we will discuss this further in the next section in the context of there being a bound-state pole dominating the amplitude. At the lighter pion mass, there is a slight tendency to a larger scattering length for amplitudes that lack an Adler zero, but the effect is barely significant. Our estimates at these two pion masses and those determined in Ref.~\cite{Briceno:2016mjc} are plotted in~\cref{fig:I0_SL_quarkmass}. An explanation of the observed behavior would be that these pion masses straddle a rapid divergence near $m_\pi \sim 300$ MeV, where the scattering length tends to $\pm \infty$ on either side of the divergence. In the next section, we will discuss how this can be related to the $\sigma$ pole undergoing a transition between Riemann sheets by passing through the $\pi\pi$ threshold.

\subsection{The $\sigma$ pole}
 \label{subsec:sigma}

The presence of singularities on the real axis below threshold can be inferred rather directly from graphs of $k \cot \delta$ against $k^2$. Since
\begin{equation}
t^I_\ell(s) = \frac{1}{\rho(s)} \frac{1}{\cot \delta^I_\ell(s) - i} = \frac{\tfrac{1}{2}\sqrt{s}}{k \cot \delta^I_\ell(s) - i k} \, ,
\end{equation}
it follows that pole singularities are present whenever the graph of $k\cot \delta$ intersects a curve, $ik = \pm \sqrt{-k^2}$ below threshold. The negative sign corresponds to a pole on the physical Riemann sheet, a \emph{bound state}, while the positive sign corresponds to the second Riemann sheet and a \emph{virtual bound state}.

\begin{figure*}[htbp]
\begin{center}
\resizebox{\textwidth}{!}{
  \raisebox{-0.5\height}{\includegraphics{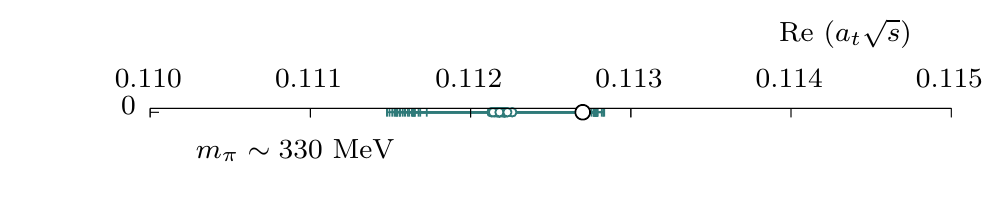}}
  \hspace{.1cm} 
  \raisebox{-0.5\height}{\includegraphics{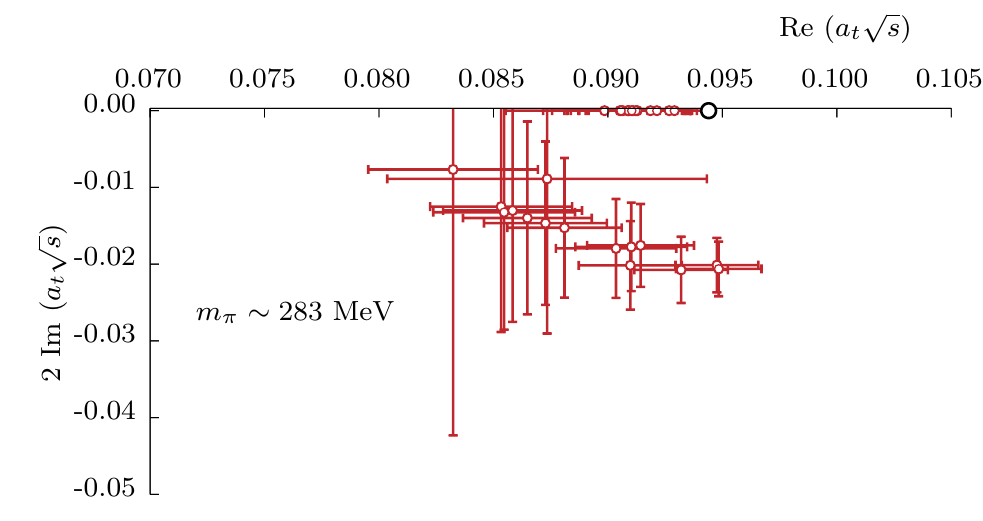}}
}
\caption{
Extracted $\sigma$ pole location for each $I=0$ $S$--wave parameterization found capable of describing the finite-volume spectra for $m_\pi\sim330$ MeV (left) and $m_\pi\sim283$ MeV (right). Left panel shows the physical sheet housing a bound-state pole, right panel shows the lower half-plane of the unphysical sheet housing either a virtual bound-state pole or a subthreshold resonance pole. For some of the parameterizations producing a virtual bound-state, a second, lighter pole is also observed on the real axis.}
\label{fig:I0Spoles}   
\end{center}
\end{figure*}

In~\cref{fig:I0Sfit} (left), for the heavier pion mass, all amplitude parameterizations intersect with $- \sqrt{-k^2}$ only slightly below threshold, with a parameterization dependence below the statistical uncertainty. This indicates the presence of a bound-state lying very close to threshold, and indeed numerical determination shows it to be statistically compatible with being at threshold, see~\cref{fig:I0Spoles}. Restricting amplitude fits to only describing levels in a small region around threshold does not change this conclusion.

As a bound-state pole approaches threshold, the value of $k\cot \delta^0_0$ at the pole location tends to $1/a^0_0$, and hence the scattering length must diverge to $-\infty$ as was suggested in the previous section. An argument due to Weinberg~\cite{Weinberg:1962hj} suggests that the scattering length ($a_0^0$), effective range ($r^0_0$), and binding energy ($\epsilon=2 m_\pi-m_\sigma$) together can be used to determine the degree to which this bound-state is of ``$\pi\pi$-molecular'' versus ``compact'' nature,

\begin{equation}
\begin{aligned}
& a^0_0=-2 \frac{1-Z}{2-Z} \frac{1}{\sqrt{m_\pi \epsilon}}\,, \quad r^0_0=-\frac{Z}{1-Z} \frac{1}{\sqrt{m_\pi \epsilon}},
\end{aligned}
\end{equation}
where $Z$ is interpreted as the probability to find the state in a compact configuration. Compatible values of $Z$ are obtained from each of these two equations suggesting that (suppressed) corrections are modest, and the resulting $Z = 0.07(4)$ suggests dominance of a $\pi\pi$ component over any compact component in the $\sigma$ at this pion mass.

In~\cref{fig:I0Sfit} (right), for the lighter pion mass, there are now two classes of parameterization. Many parameterizations capable of describing the finite-volume spectrum cross the curve $+\sqrt{-k^2}$ below threshold, indicating the presence of a virtual bound state, but some do not. Upon searching these latter amplitudes for poles off the real axis, complex conjugate pairs of poles are found off the real axis below threshold. As can be seen in~\cref{fig:I0Sfit} (right), there is not a significant qualitative difference in the amplitude above threshold between the effect of a virtual bound-state and a sub-threshold resonance. The locations of these poles are shown in~\cref{fig:I0Spoles}.

In the case of a virtual bound state lying at threshold, a similar logic to that presented above for a bound state indicates that the scattering length must diverge to $+\infty$ as the pole reaches threshold. The transition, as the pion mass increases, from scattering length diverging to $+\infty$, to reducing from $-\infty$ would therefore correspond to a pole on the second Riemann sheet moving onto the physical Riemann sheet by passing through the threshold.

\begin{figure*}[htbp]
\begin{center}
\resizebox{\textwidth}{!}{
\includegraphics{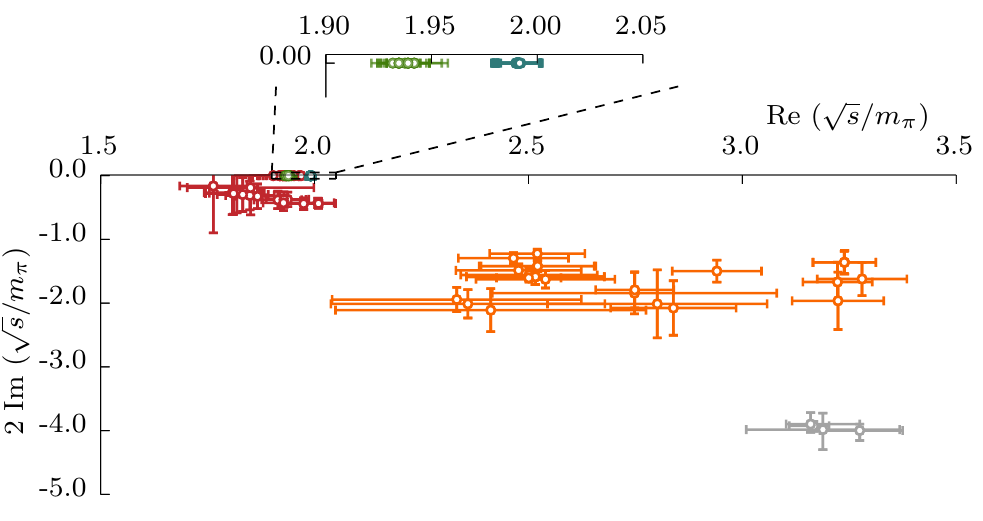}
\hspace{.1cm} 
\includegraphics{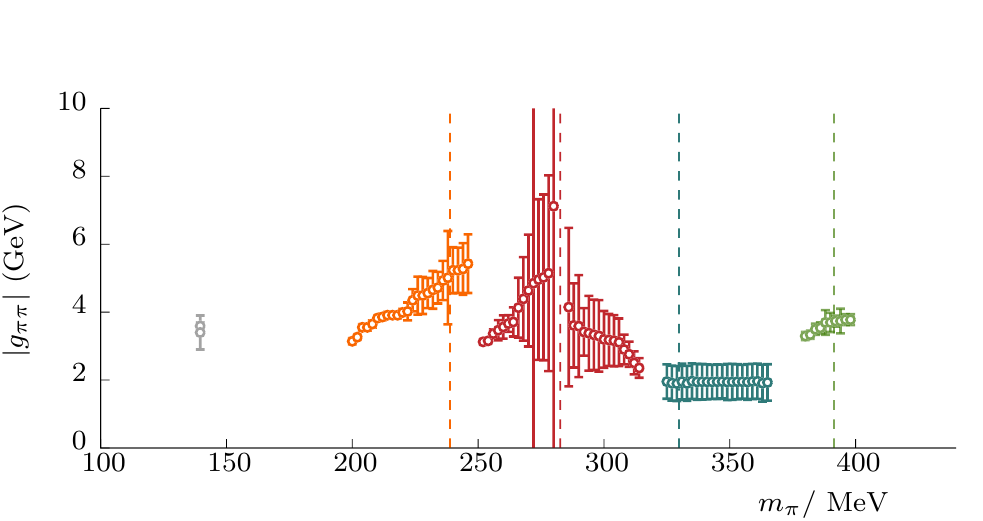}
}
\caption{
Left: $\sigma$ pole location with varying pion mass from this calculation and from calculations on lattices with the same action~\cite{Dudek:2012xn, Wilson:2015dqa}. Green ($m_\pi \sim 391$ MeV, from~\cite{Briceno:2016mjc}) and blue ($m_\pi \sim 330$ MeV) points lie on the physical sheet while red ($m_\pi \sim 283$ MeV) and orange ($m_\pi \sim 239$ MeV) points lie on the unphysical sheet (additional parameterizations have been applied to the energy levels published in~\cite{Briceno:2016mjc} to generate the spread of orange points). Right: Magnitude of the $\sigma$ pole coupling, as defined in~\cref{gNorm} at four pion masses, with different parameterizations displaced horizontally for clarity. The points at $m_\pi \sim 283$ MeV are separated into two groupings according to whether the pole in that case is a subthreshold resonance or a virtual bound-state. The dashed vertical lines locate each one of our pion masses. The result of dispersive analyses~\cite{Caprini:2005zr,Garcia-Martin:2011nna,Moussallam:2011zg} (\cite{Garcia-Martin:2011nna} only for the coupling) of experimental data is shown in gray in each plot.
}
\label{fig:sigma-quark-mass}   
\end{center}
\end{figure*}

Figure~\ref{fig:sigma-quark-mass} summarizes the $\sigma$ pole positions extracted from calculations at $m_\pi \sim 391, 330, 283$ and $239$ MeV using the same lattice action. At the heaviest two pion masses, the $\sigma$ is a stable bound-state, at $283$ MeV it is either a virtual bound-state or a subthreshold resonance (depending upon parameterization), while at $239$ MeV it appears to be a broad resonance. Qualitatively this evolution in quark mass conforms to the general scheme presented in Ref.~\cite{Hanhart:2008mx,Pelaez:2010fj,Hanhart:2014ssa} -- as the pion mass is increased from its physical value, where the $\sigma$ is a broad resonance, the complex conjugate pole pairs on the second Riemann sheet move toward the real energy axis, eventually meeting at a point below threshold. One pole then moves away towards negative infinity, becoming less relevant, while the other moves toward threshold as a virtual bound state. When this pole reaches threshold, it moves onto the physical sheet as a bound state, which becomes more deeply bound as the pion mass increases further. 

The behavior of the amplitude in the pion mass region where the conjugate pole pair meet on the real axis indicates the origin of the large statistical errors in the right panel of~\cref{fig:I0Spoles}. At this point, the derivative of the pole location with respect to amplitude parameters can diverge, leading to an inability to propagate a statistical error~\footnote{
For example, with an effective range parameterization, the dependence on the pole locations on the scattering length and effective range is given by 
$$
\frac{\partial k_R}{\partial a} = \frac{1}{a^2} \frac{1}{r k_R - i}, \;
\frac{\partial k_R}{\partial r} = -\frac{1}{2} \frac{k_R^2}{r k_R - i}, 
$$
and the location where the two poles coincide is $k_R = i/r$ with $r$ negative and with $a = -2 r$.
}. The fact that our $m_\pi \sim 283$ MeV choice appears to be close to this point generates larger statistical uncertainties on the pole position than might otherwise be expected.

The lack of a reliable determination of a \emph{second} subthreshold pole (as expected by the pole evolution argument described above) for the heavier pion mass considered in our calculation likely reflects the insensitivity of the amplitude near and above threshold (which determines the finite volume spectrum) to such a rather distant pole. Some additional constraints below threshold would be required to pin it down with certainty.

The reduction in the value of $|g_{\pi\pi}|$ observed in~\cref{fig:sigma-quark-mass} for $m_\pi \sim 330$ MeV is expected on general grounds if this point is close to the pion mass where the $\sigma$ pole passes through the threshold. Kinematic constraints on the amplitude at threshold ensure that as the pole approaches threshold, the $S$--wave coupling must behave like $g_{\pi\pi}^2 \propto \sqrt{s_R - 4 m_\pi^2}$, and hence must vanish as the pole crosses the threshold. 

As was previously observed in a lattice calculation with $m_\pi \sim 239$ MeV~\cite{Briceno:2016mjc}, the results of this paper indicate that in those cases where the $\sigma$ is unbound, even with the use of large numbers of high-precision finite-volume energy levels, the $\sigma$ pole location cannot be precisely pinned down. Different parameterizations which describe the real-energy data equally well lead to pole locations and couplings scattered well outside the statistical uncertainty, and we conclude that to reduce this systematic error it will be necessary to introduce a greater level of theoretical constraint into the determination of the scattering amplitudes. In the next section, we will describe a plausible approach to achieve this which makes use of the full set of partial-wave amplitudes across three isospins computed in this paper.

\section{Conclusions and outlook}
 \label{sec:summary}

We have reported on the extraction of $\pi\pi$ elastic scattering amplitudes across all three isospins for low partial-waves, supplementing earlier calculations on anisotropic lattices with $m_\pi \sim 391, 239$ MeV, with new calculations at two interpolating pion masses, 330 and 283 MeV. The new pion masses explore the region where we expected the $\sigma$ state appearing in the $I=0$ \mbox{$S$--wave} to transition from a bound state into a resonance. We continued the philosophy used in prior publications~\cite{Dudek:2012gj,Dudek:2012xn, Dudek:2014qha,Wilson:2014cna,Wilson:2015dqa, Dudek:2016cru, Briceno:2016mjc, Briceno:2017qmb, Woss:2018irj, Wilson:2019wfr, Woss:2019hse, Woss:2020ayi, Johnson:2020ilc} of exploring a wide range of amplitude parameterizations to test the uniqueness of the lattice QCD spectrum constrained amplitudes and their resonance content.

The isospin--2 $S$--wave at both new pion masses is found to be weak and repulsive, and we isolated a sensitivity in the extracted value of the scattering length to whether amplitude parameterizations contain a subthreshold zero like an Adler zero. The isospin--1 $P$--wave is dominated by an isolated narrow $\rho$ resonance, and we were able to establish the trajectory of the corresponding resonance pole through the complex plane as the pion mass varies. The corresponding coupling of the resonance to $\pi\pi$ was found to be essentially quark mass independent.

The isospin--0 $S$--wave, which houses the $\sigma$, shows the most dramatic change between the two pion masses considered. At $m_\pi \sim 330$ MeV the phase-shift is relatively flat over the entire region and is found to feature a bound-state $\sigma$ with a binding energy of only about \mbox{3 MeV}, while at $m_\pi \sim 283$ MeV the phase-shift rises slowly from $0^\circ$ caused by the $\sigma$ being either a virtual bound-state or a subthreshold resonance. We conclude that the $\sigma$ undergoes a transition from being a bound state to being a virtual bound state somewhere between $m_\pi \sim 283$ MeV and $m_\pi \sim 330$ MeV. 

The very different quark mass evolutions observed for the vector $\rho$ and the scalar $\sigma$ agree with the general arguments that a $P$--wave resonance can only become stable by having the complex conjugate resonance pole pair coalesce at the threshold, while an $S$--wave state need not meet this requirement. Once the pair of $S$--wave poles meet on the real axis below threshold, they evolve differently, with one of them approaching threshold as the quark mass increases. In those lattices where we find a bound $\sigma$, the pole closest to threshold determines the low energy behavior of the partial-wave.

The fact that we are unable to state with certainty whether the $\sigma$ at $m_\pi \sim 283$ MeV is a virtual bound-state or a subthreshold resonance reflects the same problem that was previously reported for the $\sigma$ at $m_\pi \sim 239$ MeV in Ref.~\cite{Briceno:2016mjc}, where equally good amplitude descriptions of the finite-volume spectrum have poles in locations scattered across the complex plane. Given the degree of systematic uncertainty associated with the choice of parameterization observed, it would not be appropriate to attempt to extrapolate the current data at unphysical pion masses to the physical pion mass.

The inability of even large numbers of high-precision lattice QCD energy levels to uniquely pin down the $\sigma$ pole location, and also to determine the location of Adler zeroes in the $I=0$ and $I=2$ $S$--wave amplitudes, are problems that most likely have a common origin. In both cases, we are required to analytically continue relatively far from where the amplitudes are constrained, which is over a limited section of the real energy axis mainly above threshold. 

We propose that a solution is to apply additional theoretical constraints to the amplitudes. In particular, the behavior of any fixed isospin partial-wave amplitude for $s<0$ is controlled by partial waves in all isospins by virtue of \emph{crossing symmetry}, and dispersion relations allow us to make practical use of this symmetry, while also ensuring good analytic properties of the amplitudes. Since we have computed amplitudes with all isospins on the same lattices in this paper, we can envisage applying a dispersion relation analysis approach to more accurately constrain partial-wave amplitudes. We are pursuing such an approach, and a publication is in preparation.

\acknowledgments
{
  We thank our colleagues within the Hadron Spectrum Collaboration, in particular D.J.~Wilson for his careful reading of the manuscript, and L.~Leskovec for useful discussions.
  
	AR, JJD and RGE acknowledge support from the U.S. Department of Energy contract DE-AC05-06OR23177, under which Jefferson Science Associates, LLC, manages and operates Jefferson Lab. 
	AR, JJD also acknowledge support from the U.S. Department of Energy award contract DE-SC0018416.

The software codes
{\tt Chroma}~\cite{Edwards:2004sx}, {\tt QUDA}~\cite{Clark:2009wm,Babich:2010mu}, {\tt QUDA-MG}~\cite{Clark:SC2016}, {\tt QPhiX}~\cite{ISC13Phi}, and {\tt QOPQDP}~\cite{Osborn:2010mb,Babich:2010qb} were used. 
The authors acknowledge support from the U.S. Department of Energy, Office of Science, Office of Advanced Scientific Computing Research and Office of Nuclear Physics, Scientific Discovery through Advanced Computing (SciDAC) program.
Also acknowledged is support from the U.S. Department of Energy Exascale Computing Project.
The contractions were performed on clusters at Jefferson Lab under the USQCD Initiative and the LQCD ARRA project. This research was supported in part under an ALCC award, and used resources of the Oak Ridge Leadership Computing Facility at the Oak Ridge National Laboratory, which is supported by the Office of Science of the U.S. Department of Energy under Contract No. DE-AC05-00OR22725.
This research is also part of the Blue Waters sustained-petascale computing project, which is supported by the National Science Foundation (awards OCI-0725070 and ACI-1238993) and the state of Illinois. Blue Waters is a joint effort of the University of Illinois at Urbana-Champaign and its National Center for Supercomputing Applications. This research used resources of the National Energy Research Scientific Computing Center (NERSC), a DOE Office of Science User Facility supported by the Office of Science of the U.S. Department of Energy under Contract No. DE-AC02-05CH11231.
The authors acknowledge the Texas Advanced Computing Center (TACC) at The University of Texas at Austin for providing HPC resources. Gauge configurations were generated using resources awarded from the U.S. Department of Energy INCITE program at Oak Ridge National Lab, and also resources awarded at NERSC. 
}


\bibliographystyle{apsrev4-1}
\bibliography{largebiblio.bib,bibi}


\end{document}